\documentclass[prb,twocolumn,showpacs,superscriptaddress]{revtex4-1}
\usepackage[english]{babel}
\usepackage[dvips]{graphicx}
\usepackage{color}
\usepackage{latexsym}
\usepackage{amsmath}
\usepackage{amsthm}
\usepackage{amsfonts}
\usepackage{amssymb}
\usepackage{romannum}

\begin{document}

\pagenumbering{arabic}

\title{Control of vibrational states by spin-polarized transport in a carbon nanotube resonator}
\author{P.\ Stadler}
\affiliation{Fachbereich Physik, Universit{\"a}t Konstanz, D-78457 Konstanz, Germany}
\author{W. Belzig}
\affiliation{Fachbereich Physik, Universit{\"a}t Konstanz, D-78457 Konstanz, Germany}
\author{G. Rastelli}
\affiliation{Fachbereich Physik, Universit{\"a}t Konstanz, D-78457 Konstanz, Germany}
\affiliation{Zukunftskolleg, Fachbereich Physik, Universit{\"a}t Konstanz, D-78457, Konstanz, Germany}
\begin{abstract}
We study spin-dependent transport in a suspended carbon nanotube quantum dot in contact with two ferromagnetic leads and with the dot's spin coupled to the flexural mechanical modes. The spin-vibration interaction induces spin-flip processes between the two energy levels of the dot. This interaction arises from the spin-orbit coupling or a magnetic field gradient. The inelastic vibration-assisted spin flips give rise to a mechanical damping and, for an applied bias voltage, to a steady nonequilibrium occupation of the harmonic oscillator. We analyze these effects as function of the energy-level separation of the dot and the magnetic polarization of the leads. Depending on the magnetic configuration and the bias-voltage polarity, 
we can strongly cool a single mode or pump energy into it. In the latter case, we find that within our approximation, the system approaches eventually a regime of mechanical instability. Furthermore, owing to the sensitivity of the electron transport to the spin orientation, we find signatures of the nanomechanical motion in the current-voltage characteristic. Hence, the vibrational state can be read out in transport measurements.
\end{abstract}
\date{\today}
\maketitle
\section{Introduction}
\label{sec:intro}
Advances in the fabrication of nanoelectromechanical systems (NEMS) \cite{Roukes:2001tm,Ekinci:2005dg}
have opened the possibility to measure extremely small forces and masses. \cite{Ekinci:2004ga,Knobel:2003kl}
As the displacements of mechanical vibrations are conveniently 
registered by electron transport measurements, NEMS may prove also useful 
technologically as ultra-sensitive detectors of charge \cite{Li:2007ex} and spin. \cite{Rugar:2004wi}
Moreover, high-frequency NEMS devices operating at cryogenic temperatures 
can themselves approach the quantum regime and pave the way for   
testing quantum mechanics in solid objects formed by a macroscopic number of 
atoms. \cite{Blencowe:2004ft,Schwab:2005vj,LaHaye:2004gs}
In fact, recent experiments already cooled a mechanical mode to its quantum ground state in different types of nanomechanical oscillators.\cite{Rocheleau:2010jd,Teufel:2012jg,SafaviNaeini:2012ih}
Furthermore, a common and promising strategy to enter the quantum mechanical regime consists in interfacing the mechanical degree of freedom with an elemental quantum object, i.e., a quantum two-level  system such as superconducting Josephson qubits \cite{Armour:2002kt}, single Andrew levels \cite{Skoldberg:2008ge,Sonne:2010jz} or single 
spins.\cite{Rabl:2009fz,Bennett:2012ha}
A successful accomplishment of this strategy was reported for a nanomechanical dilatation oscillator coupled to a phase-qubit.\cite{OConnell:2010br}
This experiment and others motivate the interest in hybrid quantum nano systems 
containing nanomechanical oscillators approaching the quantum 
regime.\cite{Rabl:2010kk,Xiang:2013hm}

Concerning spin-oscillator systems, a variety of nanomechanical devices 
have been proposed.
For instance, in magnetic resonance force microscopy experiments, a mechanical cantilever with a ferromagnetic tip can detect single spins in solid samples.\cite{Rugar:2004wi,Poggio:2010jf,Mamin:2007ff,Xue:2011kd} Alternatively, the spin can be exploited for sensing the mechanical motion as for instance in experiments with nitrogen vacancy centers. \cite{Arcizet:2011cg,Kolkowitz:2012iw} 
The interplay between mechanical motion and spin transport has been analyzed in nanomechanical torsion oscillators \cite{Mohanty:2004hx,Kovalev:2007hk,Jaafar:2009ce,Zolfagharkhani:2008ka} in which a change of the angular momentum (spin flip) of the itinerant electron creates a mechanical torque similar to the Einstein-de Haas effect. 
In another recent experiment \cite{Ganzhorn:2013fj}, the magnetization reversal of a single-molecule magnet attached to a suspended carbon nanotube (CNT)\cite{Sazonova:793522} was probed by electrical transport measurements.

Suspended carbon nanotube quantum dots (CNTQDs) \cite{Huttel:2009jd,Lassagne:2009fg,Steele:2009ko,Sapmaz:2006kv,Leturcq:2009br,Island:2012is,Laird:2012ji,Stiller:2013hm,LeRoy:2004kf,Poot:2012fh,Benyamini:2014eb} have been discussed as a suitable playground for the realization of a coherent quantum spin-vibration system. The spin of discrete electron levels on the dot can couple to the flexural vibration via an extrinsic mechanism under a magnetic field \cite{Borysenko:2008fw} or via the intrinsic spin-orbit interaction. \cite{Kuemmeth:2008hk,Rudner:2010ct,Flensberg:2010fy,Jespersen:2011cd,Palyi:2012hj} Similar mechanisms were discussed in double dots systems \cite{Ohm:2012cz,Danon:2013fq}.
Remarkably, CNTQDs play also a crucial role in spintronics.
Indeed, spin-current injection has been experimentally reported in CNTs in a spin-valve geometry with gate-field control. \cite{Tsukagoshi:1999dh,Cottet:2006iv,Jensen:2005bm,Sahoo:2005dw,Cottet:2006iv,Jensen:2005bm}  
To conclude the state of this field, we emphasize that the interplay between nanomechanical effects and spin-dependent transport can lead to interesting phenomena as mechanical self-excitations\cite{Radic:2011ie},
shuttle mechanism controlled by external magnetic field, \cite{Fedorets:2005jt} phonon lasing\cite{Khaetskii:2013jz} or cooling of mechanical vibrations. \cite{Stadler:2014hr,Bruggemann:2014gy,Atalaya:2012eb}

\begin{figure}[t!]
\begin{center}
\includegraphics[width=0.8\columnwidth,angle=0.]{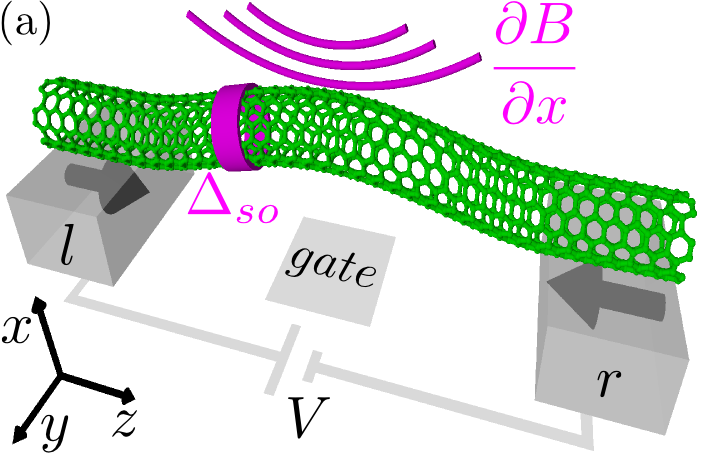}
\\ 
\vspace{0.5cm}
\includegraphics[width=0.8\columnwidth,angle=0.]{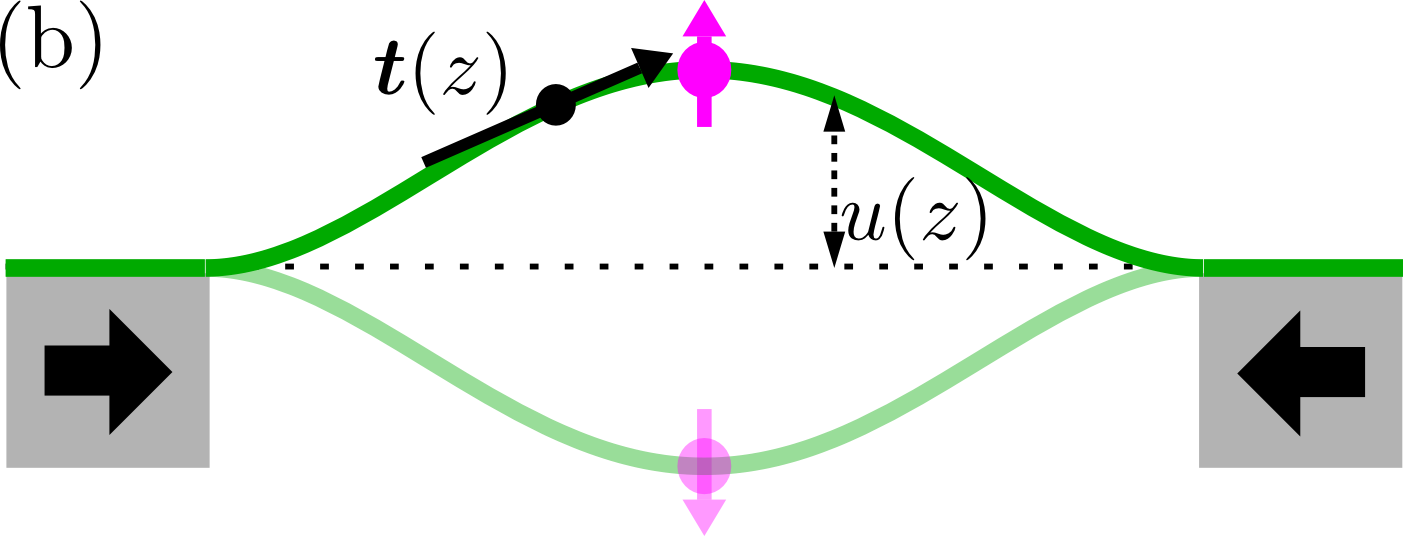}
\end{center}
\caption{(Color online) Schematic views of a carbon nanotube quantum dot suspended between two 
ferromagnetic leads. (a) The spin-vibration interaction can be either induced by
the intrinsic spin-orbit coupling $\Delta_{SO}$ or by a magnetic gradient $\partial{B}/\partial{x}$. 
(b) Due to the spin-vibration interaction, the dot spin's component $\hat{\sigma}_x$ parallel to the 
mechanical displacement $u$ couples to the flexural mode. The local tangent 
vector is denoted by $\boldsymbol{t}$.
}    
\label{fig:schema_experiment}
\end{figure}

Motivated by the growing interest in combining nanomechanics with spintronics, 
in this work we discuss the effects of the spin-vibration interaction when a suspended CNTQD 
is sandwiched  between two ferromagnets and a bias-voltage is applied. 
We consider a model with a single mechanical (flexural) mode of frequency $\omega$. 
We show that the system acts as a nanomechanical spin-valve in which spin-polarized electrons tunneling 
through the CNTQD can exchange energy with the oscillator by flipping the spin.
Such vibration-assisted spin-flip processes give rise to a mechanical damping of the oscillator 
and to inelastic transport through the CNTQD.  
Concurrently, when electric current flows through the CNTQD, the oscillator is also driven towards a steady, 
nonthermally equilibrated regime in which the average energy stored into the oscillator is 
larger (heating) or smaller (cooling) than the thermal energy. 
This corresponds to a phonon occupation different from the thermal Bose distribution at the lattice temperature. 
When the oscillator is heated by increasing the bias voltage, the damping coefficient can also vanish at a threshold voltage and then becomes negative at higher voltages. 
We obtain this result in the perturbation expansion for the spin-vibration coupling and neglecting anharmonic effects and feedback action of the resonator on the current. Such effects will eventually remove the mechanical instability. However, our results point out a special regime  of parameters in which we expect interesting effects as, for instance, vibrational lasing.
Finally, we find remarkable features in the current-voltage characteristic which are directly related 
to the non-thermal phonon occupation of the oscillator. 
Hence, transport measurements offer the possibility for monitoring the oscillator's state directly by varying the voltage polarity and/or the relative alignment of the magnetizations in the leads 
(parallel or antiparallel).

The paper is structured as follows.
In Sec.~\ref{sec:model}, we introduce the model Hamiltonian and 
derive the formulas for the mechanical damping,  the steady-state nonequilibrium phonon 
occupation, and the current using the Keldysh nonequilibrium Green's functions technique.
We calculate such quantities to the first leading order in the spin-vibration coupling strength. 
In Sec.~\ref{sec:damplingoftheoscillatorandphononoccuapation}, we discuss the nonequilibrium phonon occupation obtained 
by applying a bias voltage. 
In comparison to our previous analysis in Ref. [\onlinecite{Stadler:2014hr}], we discuss the active heating or cooling of the mechanical oscillator 
for the regimes in which (i) the system approaches a mechanical instability, (ii) a single lead is magnetically polarized. 
In Sec.~\ref{sec:current}. we discuss the effects of the spin-vibration interaction on the current.
In Sec.~\ref{sec:summary} we summarize our work.

\section{Model and Approximation}
\label{sec:model}
\subsection{Microscopic derivation of the Hamiltonian}
\label{subsec:microscopichamiltonian}
The nanomechanical spin valve that we consider consists of a suspended CNTQD in contact with ferromagnetic leads [Fig.~\ref{fig:schema_experiment}(a)]. 
In this section, we introduce the model Hamiltonian of a suspended CNTQD and derive the spin-vibration interaction induced 
by the spin-orbit coupling or by the application of a magnetic gradient. 

\setcounter{paragraph}{0}
\subsubsection{Carbon nanotube quantum dot}
In a confining potential and for vanishing magnetic field and spin-orbit interaction, the
localized electronic levels of a CNTQD are, at least, fourfold degenerate owing 
to the spin and circumferential orbital degree of freedom. \cite{Laird:2014ww}
We denote the corresponding states as $\vert \tau, \sigma \rangle$ with $\tau=\pm$ and $\sigma=\pm$ referring to the orbital and spin states, respectively. 
We choose the spin quantization axis along the $z$ direction. 
The effective low-energy Hamiltonian for a single dot shell is given by
\cite{Rudner:2010ct,Flensberg:2010fy,Palyi:2012hj}
\begin{equation}
\label{eq:H_eff_dot}
\hat{H}_{\mathrm{CNT}} =  
\frac{\Delta_{\mathrm{SO}}}{2} \hat{\tau}_3 {\bf t}(z) \cdot   \boldsymbol{ \hat{\sigma}} 
- \mu_{orb} \hat{\tau}_3  {\bf B} \cdot  {\bf t}(z) 
+  \mu_B    {\bf B}  \cdot \boldsymbol{ \hat{\sigma}} 
+
\Delta_{KK'} \hat{\tau}_1 \,,
\end{equation}
with the orbital magnetic moment $\mu_{orb}$, the Bohr magneton $\mu_B$, the intrinsic spin-orbit coupling $\Delta_{\mathrm{SO}}$, the coupling $\Delta_{KK^\prime}$ between different orbital states due to disorder, and the magnetic field $\boldsymbol{B}$. The Pauli matrices in spin (orbital) space are denoted as $ \boldsymbol{ \hat{\sigma}}=(\hat{\sigma}_x,\hat{\sigma}_y,\hat{\sigma}_z)$ ($\boldsymbol{\hat{\tau}} = [(\hat{\tau}_1,\hat{\tau}_2,\hat{\tau}_3)]$ and the local tangent vector at each point of the tube is written as ${\bf t}(z)$ whose direction varies with the position $z$ [Fig.~\ref{fig:schema_experiment}(b)]. 
The validity of the Hamiltonian (\ref{eq:H_eff_dot}) is based on the energy scale separation between the high-energy spacing associated 
to the gap due to the longitudinal and the circumferential quantization and the small coupling energies appearing in 
Eq.~\eqref{eq:H_eff_dot}. \cite{Rudner:2010ct,Flensberg:2010fy}
Moreover, since typically $\Delta_{KK'} \ll (\Delta_{SO}, \mu_{orb}B,\mu_B B)$, we neglect the coupling between different orbitals in the following as we discuss the transport far away from the regime in which the energy crossing between different orbital states occurs.

\subsubsection{Spin-vibration interaction}
The deflection associated with the flexural mode leads to a coupling of the spin
on the quantum dot with the vibration which is either mediated by the spin-orbit coupling or by a magnetic gradient. The electronic model and the coupling induced by the spin-orbit coupling were studied in Refs. [\onlinecite{Palyi:2012hj}], [\onlinecite{Ohm:2012cz}] and [\onlinecite{Danon:2013fq}].  Here, we additionally derive the coupling between the deflection and the spin due to a magnetic gradient. Such a coupling was also analyzed in Ref.~[\onlinecite{Atalaya:2012eb}].
It arises from the relative motion of the suspended 
nanotube in a magnetic gradient in addition to a homogeneous magnetic field. 
\cite{Bargatin:2003en}

We depict in Fig.~\ref{fig:schema_experiment} the choice of the coordinate axes and assume in the following that the nanotube oscillates in the $x-z$ plane. 
The deflection $\hat{u}(z)$ can be written as a linear combination of the oscillation amplitudes of the eigenmodes, 
$\hat{u}(z) =\sum_n f_n(z) u_n ( \hat{b}_n^{\phantom{g}}  + \hat{b}_n^{\dagger} )$, with the waveform $f_n(z)$, the zero-point amplitude  
$u_n={[\hbar/(m\omega_n)]}^{1/2}$, and the bosonic annihilation (creation) operators $\hat{b}$ ($\hat{b}^{\dagger}$) for a single mode with frequency $\omega_n$.
For a suspended elastic rod of length $L$,  mass line density $\rho$, and with sufficient strong tension $T$,   
the eigenfrequency is  $\omega_n=(n+1)\,\pi \sqrt{T/(\rho L^2)}$ and the waveform is given by $f_n(z)= \sqrt{2} \sin[ \pi (n+1)  z/L ]$ 
for integers $n \geq 0$. \cite{Stadler:2014hr}
Assuming that the deflections are sufficiently small, we approximate the variation of the tangent vector as $\delta{\bf t}(z)  \simeq [d\hat{u}(z)/dz,0,0]$. 
Additionally, the magnetic field along the nanotube changes by $\delta{\bf{B}}=(\partial{\bf{B}}/\partial x) \hat{u}(z)$ due to the magnetic gradient. 
Thus we expand ${\bf B} \cdot {\bf t}(z) \simeq B_z+ {\bf B} \cdot \delta{\bf t}(z)  + \delta {\bf B}   \cdot {\bf{z}}$ in which we neglect $\delta{\bf t}(z) \cdot \delta {\bf B} $ corresponding to higher-order terms in $\hat{u}$ ($\bf{z}$ denotes the unit vector in the $z$ direction).
 In the following, we assume a leading magnetic gradient $dB_x/dx$ perpendicular to the nanotube $z$ axis and neglect the variation of the $y$ and $z$ components of the magnetic field along the  $x$ axis $dB_{y,z}/dx=0$. Furthermore, we assume a vanishing magnetic field in the $x$ direction $B_x=0$. 
Inserting the expansion of $\boldsymbol{B}$ and $\boldsymbol{t}(z)$ into Eq.~\eqref{eq:H_eff_dot} we obtain\cite{Palyi:2012hj,Stadler:2014hr}
\begin{equation}
\label{eq:H_0_dot}
\hat{H}_{cnt} =  \hat{H}_{cnt}^{(0)} + \hat{H}_{SV,1}+ \hat{H}_{SV,2}  \, ,
\end{equation}
with 
\begin{align}
\hat{H}_{cnt}^{(0)} &=
\frac{\Delta_{SO}}{2}  \hat{\tau}_3\hat{\sigma}_z   - \mu_{orb}  B_z \hat{\tau}_3 +   \mu_B    B_z  \hat{\sigma}_z  \, \label{eq:H_CNT_0}  \\
 \hat{H}_{SV,1}  &=
\mu_{B}  \frac{\partial B_x}{\partial x}  \sum_n\left<  f_n(z) \right> 
u_n \left( \hat{b}_n^{\phantom{g}}  + \hat{b}_n^{\dagger} \right) \hat{\sigma}_{x} \, \\
\hat{H}_{SV,2}&=
\frac{\Delta_{SO}}{2} \sum_n\langle f_n^\prime(z) \rangle u_n \left(\hat{b}_n+\hat{b}_n^\dagger\right) \hat{\tau}_3 \hat{\sigma}_x
\, , 
\end{align}
in which the waveform $f_n$ is averaged over the electronic orbital in the dot (we also assumed that the variation of the magnetic gradient along the nanotube axis is negligible). 
For a quantum dot with symmetric orbital electronic density, the averages $\left<  f_n(z) \right> $ ($\langle f_n^\prime(z) \rangle$) vanish for all odd (even) harmonics. 
To give a simple estimation, we consider a  uniform distribution of the electronic charge on the dot. We obtain  $\left< f_0 (z)\right>  =2\sqrt{2}/\pi$ for the first even mode (the fundamental mode) and $\left< df_1 (z)/dz\right>  =2\sqrt{2}/L$ for the first odd mode.
In this way, the coupling constant $\lambda_n \simeq \mu_B  (\partial B_x/\partial x) u_n  \langle f_n(z) \rangle$ in $ \hat{H}_{SV,1}$  can be estimated as $\lambda_0=0.5$ MHz for the fundamental mode  with  
$\partial B_x/\partial x  = 5 \cdot 10^{6}$ T/m.\cite{Xue:2011kd} 
The coupling constant $\lambda_n \simeq (\Delta_{SO}/2)  u_n  \langle df_n(z)/dz\rangle$ in $\hat{H}_{SV,2}$  is estimated as $\lambda_1 \sim 2.5$ MHz for the first odd mode with $\Delta_{SO} \simeq 400$ $\mathrm{\mu eV}$. \cite{Palyi:2012hj}

\begin{figure}[t!]
\begin{center}
\includegraphics[width=0.9\columnwidth,angle=0.]{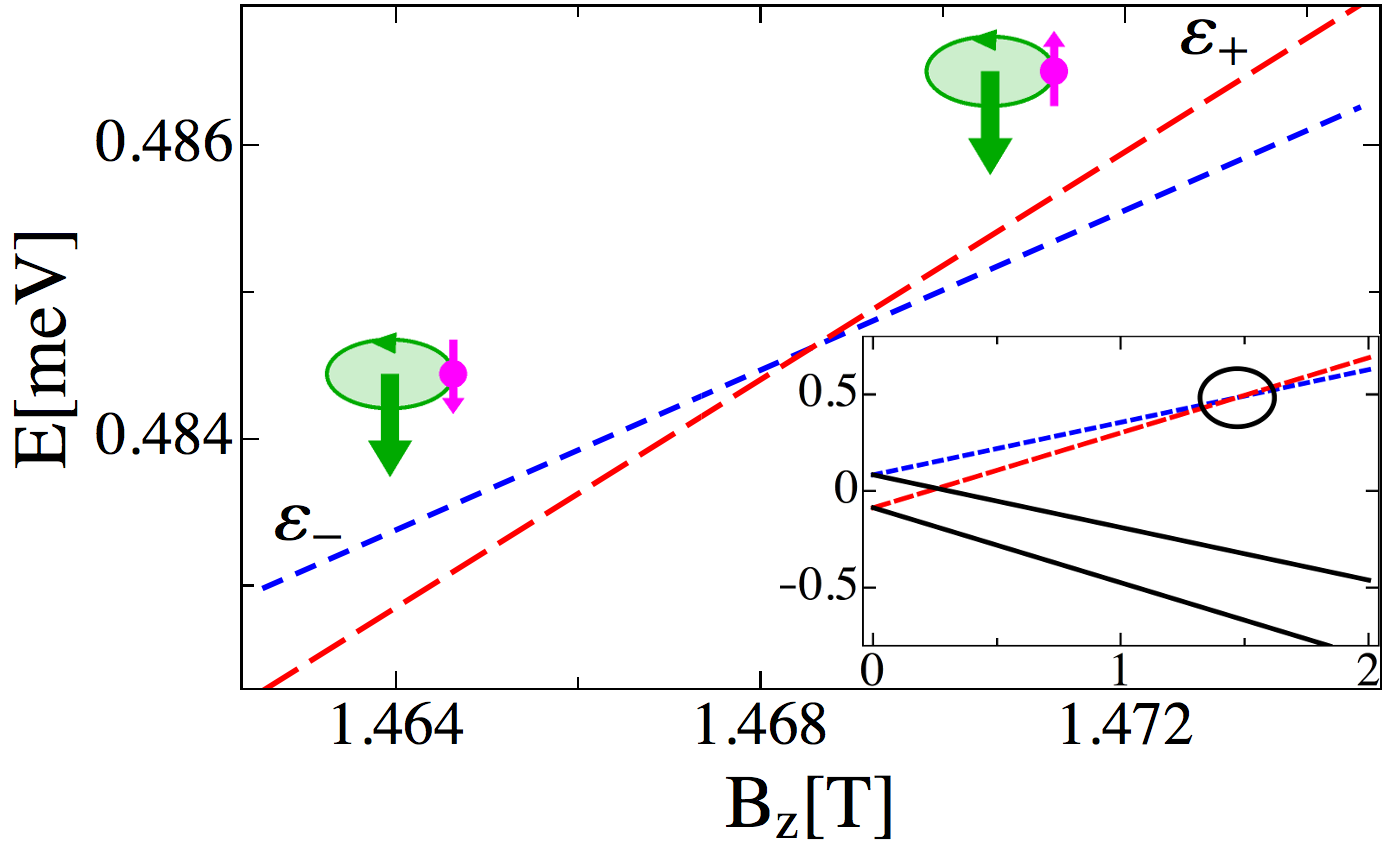}
\end{center}
\caption{(Color online) Spectrum of the Hamiltonian for a defect-free carbon nanotube quantum dot. 
The inset shows the full spectrum as a function of the magnetic field along the nanotube axis as given 
by the Hamiltonian  Eq.~(\ref{eq:H_CNT_0}). The circle in the inset points out the crossing point between the two levels reported in the main panel.
We focus on the electron transport in which only two levels of energies $\varepsilon_+$ and $\varepsilon_-$ are involved. 
They have the same orbital state and opposite spin. The sketches illustrate the direction of the orbital (large green arrow) and spin (small magenta arrow) 
magnetic moments along the $z$-axis. 
The parameters are $\Delta_{SO}=170$ $\mu$eV and $\mu_{orb}=330$ $\mu$eV/T from Ref.~[\onlinecite{Churchill:2009fy}].}    
\label{fig:CNTQD-spectrum}
\end{figure}

\subsubsection{Single mode model with two spin levels}
We now consider the suspended CNTQD embedded between
ferromagnetic leads. The leads are described by the Stoner 
model in which one assumes a spin asymmetry 
in the density of states for the spin-up and -down bands
$\rho_{\alpha\sigma}=\rho_{\alpha} (1 +\sigma p_{\alpha})$
with the degree of spin polarization in lead $\alpha$ defined as 
$p_{\alpha}=(\rho_{\alpha+} - \rho_{\alpha -})/(\rho_{\alpha +} + \rho_{\alpha -})$. 
The effect of the ferromagnets is captured by the spin-dependent
tunneling rates $\Gamma_{\alpha}^{\sigma} = \pi |t_{\alpha\sigma}|^2 \rho_{\alpha\sigma}$. 
The Hamiltonian of the whole system is given by 
\begin{equation}
\label{eq:Hamiltonian_H0}
\hat{H}
= 
\hat{H}_{l}
+
\hat{H}_{t}  
+ 
\hat{H}_{d} \, ,
\end{equation}
where the Hamiltonian for the leads $(\alpha=L,R)$ reads as
$\hat{H}_{l} = \sum_{\alpha\sigma k} \varepsilon_{k\sigma}^{\phantom{}} \hat{c}^{\dagger}_{\alpha k\sigma}  
\hat{c}_{\alpha k\sigma}^{\phantom{}}$ 
and the tunneling Hamiltonian is
$\hat{H}_{t} = \sum_{\alpha \sigma k} ( t_{\alpha\sigma}^{\phantom{}} \hat{c}^{\dagger}_{\alpha k \sigma } 
\hat{d}_{\sigma}^{\phantom{}} + h.c. )$. 
The operators $\hat{c}^{\dagger}_{\alpha k \sigma }$ ($\hat{c}_{\alpha k \sigma}^{\phantom{}}$) and 
$\hat{d}^{\dagger}$  ($\hat{d}^{\phantom{}}$) 
are the creation (annihilation) operators for the corresponding 
electronic states in the ferromagnetic leads and for the dot states. 
To discuss the effects of the spin-vibration interaction, 
we focus on a part of the spectrum of the CNTQD given by the Hamiltonian (\ref{eq:H_CNT_0}),  
i.e., the situation in which only two spin channels for the same orbital level are 
involved in the transport, as shown in  Fig.~\ref{fig:CNTQD-spectrum}.  This regime occurs when the 
orbital energy splitting is the largest energy scale in  Eq.~\eqref{eq:H_CNT_0}.

The model Hamiltonian for the two spin states of the same orbital and 
the spin-vibration interaction with a single mechanical mode of frequency $\omega$ is finally described by
\begin{equation}
\label{eq:Hamiltonian}
\hat{H}_d
= 
\sum_{\sigma} \varepsilon_{\sigma} 
\hat{d}^{\dagger}_{\sigma} \hat{d}_{\sigma}^{\phantom{}}
+ 
\lambda \hat{\sigma}_x  \left( \hat{b}^{\dagger} + \hat{b}^{\phantom{}} \right)
+
\hbar \omega \hat{b}^{\dagger}  \hat{b}^{\phantom{}}  \, ,
\end{equation}
with the energy levels $\varepsilon_\sigma = \varepsilon_0+\sigma\varepsilon_z/2$ and the splitting between the two spin states given by $\varepsilon_z$.
The $x-$component of the local spin operator in the dot   
$\hat{\sigma}_x  = \hat{d}^{\dagger}_{+} \hat{d}_{-}^{\phantom{}} 
+ \hat{d}^{\dagger}_{-} \hat{d}_{+}^{\phantom{}}$  is chosen to be perpendicular 
to the quantization axis for the spin transport. 
The index $n$ in the bosonic operators is omitted since we assume that only a single vibrational mode is relevant.

The Hamiltonian Eq.~(\ref{eq:Hamiltonian}) is similar to the well-known Anderson-Holstein 
model widely discussed in literature
\cite{Jonson:1989ts,Wingreen:1989tb,Flensberg:2003je,Braig:2003cb,Mitra:2004em,Koch:2006hu,Galperin:2006jg,Zazunov:2007hv,Galperin:2007ie,Egger:2008hh,EntinWohlman:2009iz,Maier:2011dj,Pistolesi:2009ks,Cavaliere:2010ks,Haupt:2009cu,Novotny:2011bs}
in which the quantum oscillator is linearly coupled to the dot charge 
$\hat{n}=\hat{d}^{\dagger} \hat{d}^{\phantom{}}$ of a spinless level, according to the Hamiltonian
$\hat{H}_{int} = \lambda  ( \hat{b}^{\dagger} + \hat{b}^{\phantom{}} ) \hat{n}$.
We recover such a model if the operator $\hat{\sigma}_x$ is replaced with $\hat{\sigma}_z$, i.e., when the spin-vibration interaction is parallel to the 
magnetization axis of the two leads so that the transport occurs through two spin channels separately.
The Hamiltonian Eq.~\eqref{eq:Hamiltonian} is also similar to the phenomenological model discussed in Refs.~[\onlinecite{Galperin:2009fn}] and [\onlinecite{Haupt:2010iz}] for an electron-vibration interaction invoking different dot levels. \cite{Arrachea:2014cm}
However, these previous works assumed mainly the case of non-ferromagnetic leads, 
whereas we will focus on the effects of spin-polarized tunneling on the vibration.

\subsection{Phonon Green's function}
\label{subsec:phonon propagator}
Electrons tunneling inelastically on and off the CNTQD yield a damping of the vibration with a rate $\gamma$ and a frequency renormalization $\Delta\omega$.
Moreover, an electron current flowing through the CNTQD drive the oscillator to a non-thermal state with phonon occupation $n \neq n_B(\omega)$ [$n_B(\omega)$ is the Bose distribution] 
if  the intrinsic coupling of the oscillator to the external thermal bath is sufficiently small. 
To address these effects, we use the Keldysh nonequilibrium Green's functions technique.

We start with the Dyson equation for the phonon Green's function in Keldysh space defined as 
\begin{equation}
 \label{KeldyshPhononProp}
\check{D}(\varepsilon) =  \check{d}(\varepsilon)  +  \check{d}(\varepsilon)\left[\check{\Pi}(\varepsilon)+\check{\Sigma}_{0}(\varepsilon)\right] \check{D}(\varepsilon) \, , 
\end{equation}
in which the retarded and Keldysh Green's functions are defined as 
$D^R(t)  = -i \theta(t) \langle  [ \hat{A}(t),   \hat{A}(0) ] \rangle>$
and
$D^K(t)  = -i  \langle  \{\hat{A}(0), \hat{A}(t) \} \rangle $
with $\hat{A}(t) = \hat{b}^\dagger(t)+\hat{b}(t)$ and the commutator (anti-commutator) 
$[\, , \, ]$ ($\{\, , \, \}$). We used the triangular Larkin-Ovchinnikov representation
\begin{equation}
\check{D}(t) = 
\begin{pmatrix} D^R(t) & D^K(t) \\ 0 & D^A(t) \end{pmatrix}.
\end{equation}
and 
we set $\hbar=k_B=1$. The bare phonon Green's functions in Eq.~\eqref{KeldyshPhononProp} are given by 
(using $\eta$ as an infinitesimal small real part)
\begin{align}
d^{R,A}(\varepsilon)&=2 \omega/[(\varepsilon\pm i\eta)^2+\omega^2], \\
d^{K}(\varepsilon)&=-2\pi i \left[\delta(\varepsilon\mathord-\omega)\mathord+\delta(\varepsilon\mathord+\omega)\right] \mathrm{coth}[\omega/(2T)] \, .
\end{align}
In Eq.~(\ref{KeldyshPhononProp}), 
$\check{\Pi}(\varepsilon)$ corresponds to the phonon self energy (polarization diagram) associated to the spin-vibration interaction 
between the oscillator and the electrons [see Fig.~\ref{fig:self-energy}(a)].  
To the leading order in the coupling strength of the spin-vibration interaction,  
the three components of the phonon self-energies are given by:
\begin{align}
\Pi^{R/A}_{}(\varepsilon) = -i\frac{\lambda^2}{2}\sum_\sigma \left[ 
{G}_{-\sigma}^{K}(\varepsilon^\prime) 
\circ {G}_{\sigma}^{A/R}(\varepsilon^\prime-\varepsilon)  \right. \nonumber \\ \left.
+{G}_{-\sigma}^{R/A}(\varepsilon^\prime) \circ {G}_{\sigma}^{K}(\varepsilon^\prime-\varepsilon) \right] \, ,  \label{eq:pi_R}  
\end{align}
\begin{align}
{\Pi}^K_{}(\varepsilon) =
-i\frac{\lambda^2}{2}\sum_{\sigma} \left[
{G}_{-{\sigma}}^K(\varepsilon^\prime) \circ {G}_{\sigma}^K(\varepsilon^\prime-\varepsilon)  
 \right. \nonumber \\ \left. 
+{G}_{-{\sigma}}^R(\varepsilon^\prime)\circ {G}_{\sigma}^A(\varepsilon^\prime-\varepsilon)+ 
{G}_{-{\sigma}}^A(\varepsilon^\prime) \circ {G}_{\sigma}^R(\varepsilon^\prime-\varepsilon)
\right]  \, . \label{eq:pi_K}
\end{align}
The symbol $\circ$ denotes the convolution product $a(x) \circ b(x-y) = \int_{-\infty}^\infty  (dx/2\pi) a(x)b(x-y)$.
Note that the interaction vertex due to the spin-vibration couples 
only spins of opposite direction [see Fig.~\ref{fig:self-energy}(a)].  
The electron Green's functions of the 
dot appearing in Eqs.~\eqref{eq:pi_R} and \eqref{eq:pi_K} 
are those associated to the Hamiltonian with vanishing spin-vibration interaction. 
They correspond to the exactly solvable problem of two dot levels coupled to the leads and they are given by 
\begin{align}
{G}_\sigma^{R,A}(\varepsilon) &= (\varepsilon-\varepsilon_\sigma\pm i\Gamma_l^\sigma\pm i\Gamma_r^\sigma)^{-1} \, ,\label{eq:dotgreensfunction_1}  \\
{G}_\sigma^K(\varepsilon) &= -2i{G}_\sigma^R(\varepsilon) \{\Gamma_l^\sigma [1\mathord-2f_l(\varepsilon)]\mathord+\Gamma_r^\sigma [1\mathord-2f_r(\varepsilon)]\} {G}_\sigma^A(\varepsilon)\nonumber\\
&= -2i\frac{\Gamma_l^\sigma (1\mathord-2f_l(\varepsilon))\mathord+\Gamma_r^\sigma (1\mathord-2f_r(\varepsilon))}{
(\varepsilon-\varepsilon_\sigma)^2+(\Gamma_l^\sigma+\Gamma_r^\sigma)^2}\, .  \label{eq:dotgreensfunction_2} 
\end{align}
Here, the Fermi functions of the leads are denoted by     
$f_{\alpha}(\varepsilon) \mathord= {\left\{ 1 + \exp\left[ (\varepsilon-\mu_{\alpha}) /T \right] \right\}}^{-1}$ 
with the lead chemical potentials $\mu_{\alpha}$ and $\alpha=l,r$. We also employed the wide band approximation by neglecting the energy dependence of the coupling rates $\Gamma_{r/l}^\sigma$.

To take into account the intrinsic damping of the oscillator, we additionally include a
self-energy $\check{\Sigma}_{0}(\varepsilon)$ in the phonon Dyson equation Eq.~(\ref{KeldyshPhononProp}). 
Such a self-energy can be calculated by assuming that the environment is formed by a bath of independent harmonic oscillators
(Caldeira-Leggett model) with a low-frequency linear dispersion for the spectral function (see Appendix \ref{Appendix:Phononenvironment} for further details). 
From this phenomenological model, one obtains the expressions  
\begin{eqnarray}
\mbox{Im }{\Sigma}_{0}^R(\varepsilon) &=& -\varepsilon/Q \label{eq:S_0R} \, ,\\    
{\Sigma}_{0}^K(\varepsilon) &=& -2i\varepsilon\mathrm{coth}(\varepsilon)/Q  \, ,\label{eq:S_0K}
\end{eqnarray}
in which the coefficient $Q$ corresponds to the quality factor of the oscillator. 

Finally, we obtain the phonon Green's function by solving the Dyson equation \eqref{KeldyshPhononProp},
\begin{align}
{D}^{R}(\varepsilon) 
&=
\frac{2\omega}{\varepsilon^2-\omega^2-2\omega[{\Pi}^{R}_{}(\varepsilon)+{\Sigma}_0^R(\varepsilon)]} \nonumber\\ 
&\simeq
\frac{1}{\varepsilon-\tilde{\omega}+i\gamma_{tot}}-\frac{1}{\varepsilon+\tilde{\omega}+i\gamma_{tot}} \, , \label{eq:D_RA} \\
{D}^K(\varepsilon) &={D}^R(\varepsilon)  [\Pi^K_{}(\varepsilon)+\Sigma_0^K(\varepsilon)] {D}^A(\varepsilon) \nonumber \\ 
&\simeq 
[\Pi^K_{}(\varepsilon)+\Sigma_0^K(\varepsilon)] \sum_{s=\pm}
\frac{1}{ {(\varepsilon+s \tilde{\omega})}^2 + \gamma_{tot}^2 } \label{eq:D_K} \, .
\end{align}
We introduced the renormalized frequency $\tilde{\omega}= \omega+\Delta\omega$ with 
 $\Delta \omega= \mbox{Re} [{\Pi}^R(\omega)+\Sigma_0^R]$.\cite{Kaasbjerg:2013jt} 
In the following, we set $\tilde{\omega}\rightarrow \omega$. 
In the approximations in Eqs.~(\ref{eq:D_RA}) and (\ref{eq:D_K}) we expanded 
the self-energies and the retarded phonon Green's functions around $\varepsilon \simeq \omega$.
Furthermore, we also introduced  the total mechanical damping coefficient 
as $\gamma_{tot}(\omega)=- \mbox{Im} [{\Pi}^R(\omega)+{\Sigma}_0^R(\omega)]$.

The total mechanical damping can be also written as $\gamma_{tot}= \gamma_0 + \gamma$ 
with the intrinsic damping coefficient  $\gamma_0=-\mbox{Im}\,\Sigma^R_0(\omega)=\omega/Q$ of the oscillator 
and the damping $\gamma=-\mbox{Im}\,\Pi^R(\omega)$ associated to the interaction with the electrons. 
We assumed the underdamped regime for the mechanical oscillator $\gamma,\gamma_0 \ll \omega$ 
which further justifies the approximation of the self energy $\check{\Pi}(\varepsilon)$ to the leading order 
in the spin-vibration coupling strength.
Using Eqs.~(\ref{eq:pi_R}),(\ref{eq:dotgreensfunction_1}), and (\ref{eq:dotgreensfunction_2}), and after some algebra, the explicit form for the damping coefficient reads as
\begin{eqnarray}
\gamma &=&  \sum_{\alpha,\beta=l,r} \sum_{s=\pm} \, s \, \gamma_{\alpha\beta}^s \, , \label{eq:main_gamma} \\
\gamma^{s}_{\alpha\beta}  &=& 
\frac{\lambda^2}{2}   \int \frac{d\varepsilon}{2\pi} 
T_{\alpha\beta}^s(\varepsilon,\omega)
 f_{\alpha}(\varepsilon)\left[1\mathord- f_{\beta}( \varepsilon+ s\omega)\right] \, , \label{eq:rates}
\end{eqnarray}
with  the functions  
 \begin{eqnarray}
 T_{\alpha\beta}^s(\varepsilon,\omega) &=& 4\sum_{\sigma} \Gamma_{\alpha}^{\sigma} \Gamma_{\beta}^{-\sigma} \vert G_{\sigma}^R(\varepsilon) \vert^2 \vert G^R_{-\sigma}(\varepsilon+s\omega) \vert^2 . \label{eq:T_ab}
\end{eqnarray}  
The coefficients $\gamma^{s}_{\alpha\beta}$ correspond to the rates for vibration-assisted inelastic processes in which a spin flip occurs for one electron tunneling from lead $\alpha$ to lead $\beta$ accompanied by the absorption ($s =+$) or emission ($s=-$) of a vibrational energy quantum $\omega$.  
Equation \eqref{eq:main_gamma} also shows that the damping $\gamma$ is given by a sum of positive and negative terms associated to the
processes of emission and absorption of energy, respectively. 
From this observation we can anticipate that, contrary to the intrinsic damping induced by the environment for which we assume $\gamma_0>0$, 
the oscillator can approach a regime in which $\gamma < 0$ for certain parameter ranges when the phonon emission processes overcome the phonon absorption processes. 
In other words, the system reaches a threshold at which the total damping coefficient vanishes $\gamma_{tot}=0$.
Beyond this point, one obtains the result  $\gamma_{tot}<0$ pointing out a mechanical instability region. 

Applying a bias voltage, the electron current drives the oscillator towards a nonequilibrium steady state  with an occupation 
$\langle \hat{b}^{\dagger} \hat{b} \rangle = \bar{n}= [(i/8\pi) \int d\varepsilon D^K(\varepsilon)]- 1/2$.
In the limit $\gamma_{tot} \ll (\omega,\Gamma_l,\Gamma_r,T,eV)$ and
separating the contributions of the intrinsic damping and of the spin-vibration interaction, the occupation can be written as
\begin{equation}
\label{eq:n_bar}
\bar{n}=\frac{\gamma_0 n_B(\omega)+\gamma n}{\gamma_0+\gamma}.
\end{equation}
Hence, the steady-state phonon occupation is the result of the competition between 
the interaction of the mechanical oscillator with the thermal bath and the interaction with the tunneling electrons.
Using Eqs.~\eqref{eq:pi_K}, \eqref{eq:dotgreensfunction_1}, \eqref{eq:dotgreensfunction_2}, and (\ref{eq:D_K}), 
the expression for the electronic contribution to the average occupation induced by the spin-vibration interaction reads as
\begin{equation}
\label{eq:main_results}
n = \frac{1}{\gamma}\sum_{\alpha\beta s} s \gamma^{s}_{\alpha\beta} n_B[\omega+s(\mu_{\alpha}-\mu_{\beta})]\,. 
\end{equation}
With the notation $\delta\gamma_{\alpha\beta}=\gamma_{\alpha\beta}^+-\gamma_{\alpha\beta}^-$, Eq.  \eqref{eq:n_bar} can be written as
\begin{widetext}
\begin{equation}
\bar{n}=\frac{(\gamma_0+\delta\gamma_{ll}+\delta\gamma_{rr})n_B(\omega)+(\gamma_{lr}^+-\gamma_{rl}^-)n_B(\omega+eV)+ (\gamma_{rl}^{+}-\gamma_{lr}^{-})n_B(\omega-eV) }
{\gamma_0+\delta\gamma_{ll}+\delta\gamma_{rr}+\delta\gamma_{lr}+\delta\gamma_{rl}} \,.
\label{eq:n_bar2}  
\end{equation}
\end{widetext}
The inelastic spin-flip processes involving only a single lead with rates $\delta\gamma_{rr}$ and $\delta\gamma_{ll}$ correspond to electrons tunneling on the dot, flipping the spin by exchanging a vibrational energy quantum, and 
then coming back to the initial lead.  Since the two leads have the same temperature as the external thermal bath, such processes drive the phonon occupation towards the equilibrium occupation $n_B(\omega)$, as it is shown in 
Eq. \eqref{eq:n_bar2}.

\subsection{Lowest order perturbation theory for the current}
\label{subsec:electronicgreensfunctions}
The transport properties through the system 
are calculated using the same technique, viz., the Keldysh-Green's functions. 
To understand the effect of the spin-vibration interaction, in this work 
we calculate the correction to the current to the first leading order in the spin-vibration coupling. \cite{Mitra:2004em,Egger:2008hh,Rammer:2007,Cuevas-Scheer:2010,Walter:2011kw}
The current through the left contact can be expressed as ($e>0$)
\begin{equation}
\label{eq:I_current}
I_l =  -e \langle \frac{d\hat{N}_l}{dt} \rangle =
\frac{2 e}{h} \,\, 
\mbox{Re}\left[
\sum_{k \sigma}  t_{l\sigma}
\int^{+\infty}_{-\infty} \!\!\!\!\!\!\!\! d\varepsilon \,\, 
\mathcal{G}^{<}_{d\sigma,lk\sigma}\!\! \left( \varepsilon \right)
\right]
\, ,
\end{equation}
in which  $\langle \dots \rangle$ denotes the standard quantum statistical 
average and $\mathcal{G}^<_{d\sigma,lk\sigma}(\varepsilon)$ 
the Fourier transform of the lesser Green's function 
$\mathcal{G}^<_{d\sigma,lk\sigma}(t,t^\prime) =i \langle \hat{c}_{lk\sigma}^{\phantom{}} (t^\prime) \hat{d}_\sigma^{\dagger}(t)  \rangle$.\cite{Mahan:2000,Bruus-Flensberg:2004} 
The corresponding Green's function on the Keldysh contour is defined as
$\mathcal{G}_{d\sigma,lk\sigma}(\tau,\tau^\prime) =-i \langle T_{c} 
\hat{c}_{lk\sigma}^{\phantom{}} (\tau) \hat{d}_{\sigma}^{\dagger}(\tau^\prime)  \rangle$ 
with the time-ordering operator $T_{c}$ along the Keldysh contour.
Transforming from the contour variable $\tau$ to the real time and
using the Larkin-Ovchinnikov rotation, we introduce the triangular matrix
representation $\check{\mathcal{G}}$ such that $\check{\mathcal{G}}$ has the  
three components $\mathcal{G}^{R,A,K}$.
From standard diagrammatics we obtain the Dyson equation 
$\check{\mathcal{G}}_{d\sigma,l\sigma} = 
\check{\mathcal{G}}_{d\sigma,d\sigma} \check{t}^*_{l\sigma} 
\check{g}_{lk\sigma}$ 
where $\check{g}_{lk\sigma}$ denotes the Keldysh Green's function 
for vanishing tunneling and spin-vibration interaction.
Inserting the lesser element $\mathcal{G}_{d\sigma,lk\sigma}^<=(\mathcal{G}_{d\sigma,lk\sigma}^K-\mathcal{G}_{d\sigma,lk\sigma}^R-\mathcal{G}_{d\sigma,lk\sigma}^A)/2$ of $\check{\mathcal{G}}_{d\sigma,lk\sigma}$ in the current \eqref{eq:I_current} one obtains
\begin{equation}
\label{eq:I_current_2} 
I_l\mathord=  
\frac{e}{h}\sum_{\sigma}  \Gamma_{l}^{\sigma}  
\mathrm{Im}\!\!
\int \!\!  d\varepsilon
\{
2[1\mathord-2f_l(\varepsilon)] \mathcal{G}^{R}_{d\sigma,d\sigma}(\varepsilon)
-\mathcal{G}^{K}_{d\sigma,d\sigma}(\varepsilon)
\}.
\end{equation}
The problem then reduces to the calculation of the dot-dot Green's functions $\mathcal{G}^{K,R,A}_{\sigma\sigma}$ (neglecting the index $dd$).
We expand the Green's function on the Keldysh contour
$\mathcal{G}_{\sigma\sigma}(\tau,\tau^\prime)=-i\langle T_c \hat{d}_{\sigma}(\tau)\hat{d}_{\sigma}^\dagger(\tau^\prime)\rangle$ to the order $\lambda^2$ treating the spin-vibration interaction as the perturbation. 
Finally, we transform the contour variable to the real time and use the Larkin-Ovchinnikov transformation
to represent the perturbation expansion in frequency space as
\begin{equation}
\label{eq:checkG}
\check{\mathcal{G}}_{\sigma\sigma}(\varepsilon) = 
\check{G}_{\sigma}(\varepsilon)+\check{G}_{\sigma}(\varepsilon)
\check{\Sigma}_{-\sigma-\sigma}(\varepsilon)\check{G}_{\sigma}(\varepsilon) \, .
\end{equation}
The corrections to the current are obtained by inserting the retarded and the Keldysh 
element of the perturbative expansion \eqref{eq:checkG}
into Eq.~\eqref{eq:I_current_2}.

The elements of the self energies $\check{\Sigma}_{\sigma\sigma}$ due to the 
spin-vibration interaction in  Eq.~\eqref{eq:checkG} are denoted as 
${\Sigma}_{\sigma\sigma}^{R,A,K}$ and they are given by 
\begin{align}
{\Sigma}_{\sigma\sigma}^{R,A}(\varepsilon)  &\mathord=i\frac{\lambda^2}{2}[ {D}^{R,A}(\varepsilon^\prime) \mathord\circ {G}_{\sigma}^{K}(\varepsilon\mathord-\varepsilon^\prime) \mathord+  {D}^{K}(\varepsilon^\prime) \mathord\circ {G}_{\sigma}^{R,A}(\varepsilon\mathord-\varepsilon^\prime)]  	\label{eq:Sigma_RA}, \\
{\Sigma}_{\sigma\sigma}^{K}(\varepsilon)    &\mathord= i\frac{\lambda^2}{2}\sum_{\zeta=R,A,K} {D}^{\zeta}(\varepsilon^\prime) \circ {G}_{\sigma}^{\zeta}(\varepsilon-\varepsilon^\prime)  	\label{eq:Sigma_K}
\end{align}
with the phonon Green's functions ${D}^{R,A,K}$ of Eqs.~\eqref{eq:D_RA} and \eqref{eq:D_K}.

\begin{figure}[t!]
\begin{center}
\includegraphics[width=0.42\columnwidth]{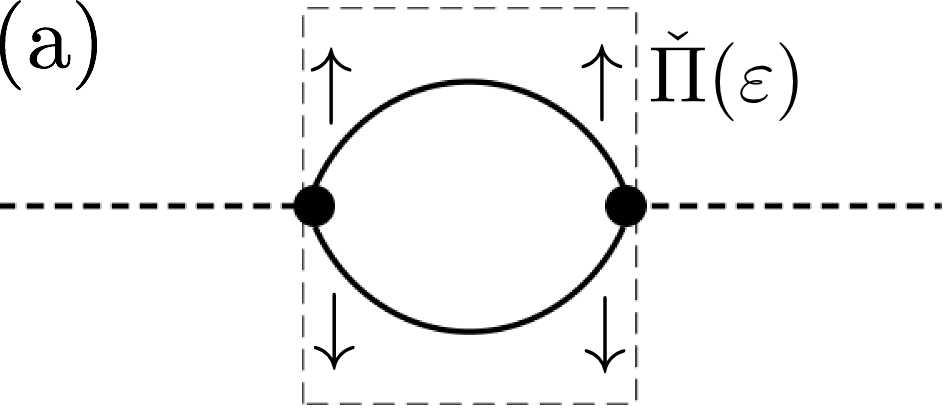} \hspace{0.5cm}
\includegraphics[width=0.42\columnwidth]{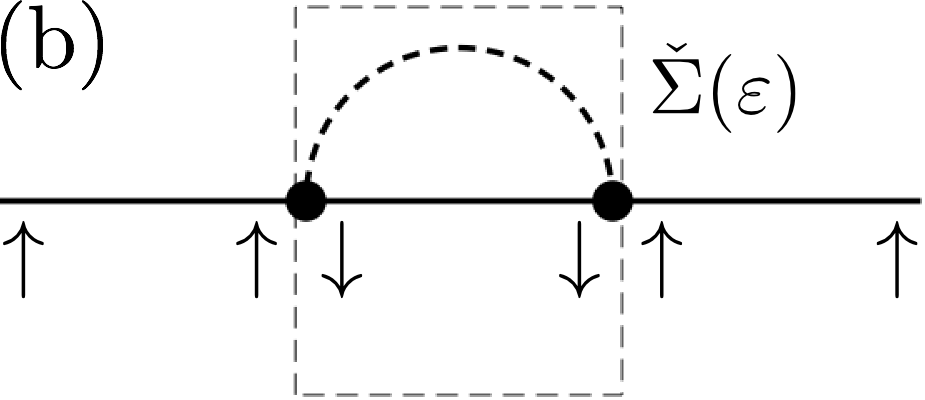}
\end{center}
\caption{
Leading-order diagrams corresponding to the perturbation expansion of the phonon Green's function $\check{D}(\varepsilon)$ (a) and the electronic Green's function $\check{\mathcal{G}}(\varepsilon)$ (b). The plane lines indicate the electronic Green's function $\check{G}(\varepsilon)$ for vanishing spin-vibration interaction. 
The dashed lines represent the bare phonon Green's functions. 
The filled circle is the vertex for the spin-vibration interaction with coupling constant $\lambda$ 
which couples electronic Green's functions of opposite spin with a phonon Green's function.
}
\label{fig:self-energy}
\end{figure}

If we compare our model with the Anderson-Holstein model, we observe that 
for the spin-vibration interaction here discussed, the tadpole diagram vanishes 
as the interaction vertex connects electron Green's functions with opposite spin [ 
see Fig.~\ref{fig:self-energy}(b)].
Hence, the expression in Eqs.~\eqref{eq:Sigma_RA} and \eqref{eq:Sigma_K} for  
the rainbow diagram represents the only finite contribution to the self-energy to the leading order. 
On the other hand, the self-energy itself $\check{\Sigma}^{}_{\sigma\sigma}$ is similar to the 
analytic expression for the Anderson-Holstein model,\cite{Mitra:2004em,Egger:2008hh,Rammer:2007}, 
except the spin dependence due to the spin-dependent interaction
(see Appendix \ref{Appendix:retardedselfenergy} for further details).

\section{Damping of the oscillator and phonon occupation}
\label{sec:damplingoftheoscillatorandphononoccuapation}

An applied voltage drives the oscillator to a nonequilibrium state with phonon occupation $\bar{n}$.
This non thermal occupation strongly depends on the configuration of the lateral ferromagnets (parallel or antiparallel magnetization configuration).

In Sec.~\ref{subsec:instability}, we discuss the state of the mechanical oscillator for the antiparallel configuration.
In a previous work Ref.~[\onlinecite{Stadler:2014hr}], we found that the antiparallel configuration allows for quantum ground-state cooling even at finite polarization of the leads. 
Here, we focus on the strong heating of the oscillator which is the precursor of a mechanical instability.
Such a regime is equivalent to a region in which phonon lasing was recently discussed for another   
spin-valve system. \cite{Khaetskii:2013jz}
In Sec.~\ref{subsec:singlepol}, we discuss the results for active cooling of the oscillator 
with a single polarized lead. 
The results of the phonon occupation in the parallel magnetization configuration are briefly 
summarized in Sec.~\ref{subsec:parallelconfiguration}. 

\subsection{Strong heating and mechanical instability}
\label{subsec:instability}

For strong enough driving, the system can approach a mechanical instability when the total damping rate vanishes $\gamma_{tot}= 0$. 
To gain an insight into the problem, we consider the state of the mechanical oscillator for fully polarized ferromagnets ($p_r=-p_l=p=1$). 
This assumption simplifies the discussion as the single lead spin-flip processes vanish $(\delta\gamma_{ll}=\delta\gamma_{rr}=0)$.
In this limit, the expression for the phonon occupation Eq.~(\ref{eq:main_results}) reads as
\begin{equation}
n_{(p=1)} \!\! = \!\!
\frac{ (\gamma_{lr}^+-\gamma_{rl}^-)n_B(\omega+eV)+ (\gamma_{rl}^{+}-\gamma_{lr}^{-})n_B(\omega-eV) }{\delta\gamma_{lr}+\delta\gamma_{rl}} .
\end{equation}
The formula for the mechanical damping Eq.~(\ref{eq:main_gamma}) reduces to $\gamma = \delta\gamma_{lr}+\delta\gamma_{rl}$  
and the total sign of the damping coefficient is now determined by the competition between the absorption and emission  processes.
Furthermore, we can consider the high-voltage approximation  $|eV| \gg (T,\omega$) in which, for instance, electrons tunneling from the right to the left lead 
are Pauli blocked for positive applied voltage  and we can neglect the corresponding rate $\gamma_{rl}^{s} \ll \gamma_{lr}^{s}$. 
The mechanical damping reduces to $\gamma\simeq\delta \gamma_{lr}$ for $eV>0$ and  $\gamma\simeq\delta \gamma_{rl}$ for $eV<0$.
Similarly, the phonon occupation reads as
\begin{eqnarray}
n_{(p=1)} &\simeq&n_{(p=1)}^{(+)} ={\left( \gamma_{lr}^{+}/ \gamma_{lr}^{-}  -1  \right)}^{-1} \quad (eV>0) \label{eq:n_p1_plus} \\
n_{(p=1)} &\simeq&n_{(p=1)}^{(-)} = {\left( \gamma_{rl}^{+} / \gamma_{rl}^{-}  -1 \right)}^{-1} \quad (eV<0) \label{eq:n_p1_minus} \, .
\end{eqnarray}
Equations \eqref{eq:n_p1_plus} and \eqref{eq:n_p1_minus} show that the phonon occupation is determined by the ratio between the absorption and emission rates.
In particular, strong heating $(n\gg1)$ is expected  if the mechanical damping coefficient vanishes $\delta \gamma_{\alpha\beta} \rightarrow 0^+$.

To understand the behavior of these rates, it is useful, as a first step, to approximate the rates for relatively large 
energy separation $\varepsilon_z$ such that mainly either the spin-up or spin-down level is involved in transport.  
For this reason, we first discuss the phonon occupation for $\varepsilon_z\gg \omega$ without intrinsic damping $(\gamma_0 = 0)$ and, second, 
we focus on the resonant case $\varepsilon_z=\omega$ including also the intrinsic damping $(\gamma_0 > 0)$. 

\subsubsection{Single level regime}

\begin{figure}[t!]
\begin{center}
\includegraphics[width=0.9\columnwidth]{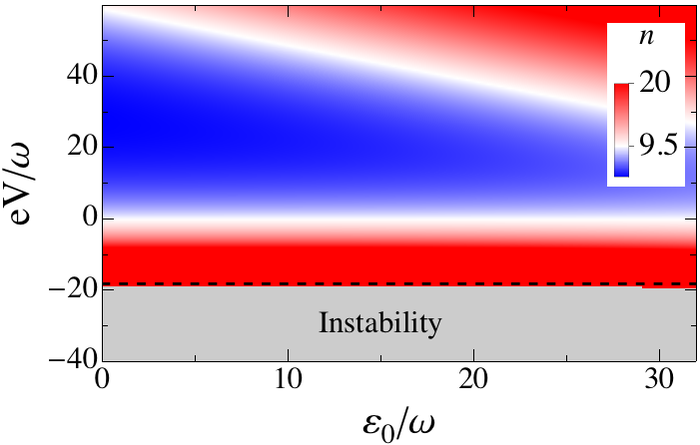}
\end{center}
\caption{(Color online) Phonon occupation $n$ as function of  the bias voltage $eV$ and gate voltage $\varepsilon_0$. 
The parameters are $p_l=-1$ and $p_r=1$, $\Gamma^{}_l = \Gamma^{}_r = 0.2 \omega$,
and $T= 10 \omega$. White color corresponds to $n_B(\omega)\approx 9.5$.  
Here we assume a vanishing external damping $\gamma_0=0$, a large spin splitting $\varepsilon_z = 10 T=100\omega$, and the chemical potential fixed to $\mu_r = \varepsilon_0 - eV$ and $\mu_l=\varepsilon_0$.
The instability regions (in gray) correspond to $\gamma<0$ and the
dashed (black) line correspond to the analytical formula for the threshold $\gamma=0$ (see text).}
\label{fig:instability}
\end{figure}

The phonon occupation for $\varepsilon_z=10 T$ and vanishing external damping is reported in Fig.~\ref{fig:instability}. 
We observe that for $eV>0$ the oscillator can be cooled or heated, whereas for $eV<0$
 the oscillator is strongly heated by increasing the bias voltage. 
 The region $eV>0$ was discussed in a previous work\cite{Stadler:2014hr} and hereafter we focus on $eV<0$.
 Beyond a certain threshold $-eV\gtrsim 20\omega$ the system approaches a mechanical unstable region.
This threshold is given by a vanishing damping $\gamma=0$. Approaching the 
threshold $\gamma=0$ from the stable region $ \gamma>0$, we have that the oscillator is strongly 
overheated with $n \gg n_B(\omega)$ since the mechanical oscillator is almost undamped 
for $\gamma \agt 0$ and it can store a large amount of energy.

\begin{figure}[t!]
\begin{center}
\includegraphics[width=1\columnwidth]{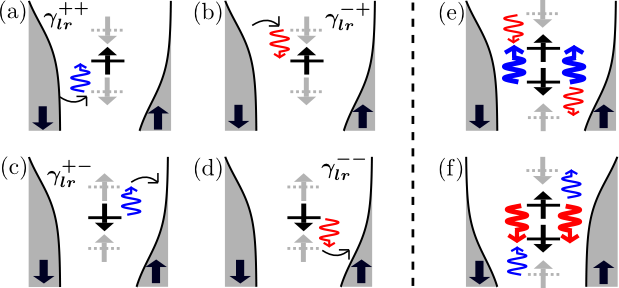}
\end{center}
\caption{(Color online) Schematic picture of the energy levels, Fermi functions, and the spin-flip processes with rate $\gamma_{lr}^{s\sigma}$ for fully polarized ferromagnets. In (a)-(d), a single level contributes to the inelastic transport which is characterized by the absorption (upwards blue arrows) or the emission (downward red arrows) of a vibrational energy quantum. In (e) and (f), the resonant condition $\varepsilon_z=\omega$ is fulfilled. When the transport is mainly determined by the process shown in (e),  optimal ground-state cooling of the oscillator is achieved. 
On the contrary, when the transport is dominated by the process shown in (f), a strong heating occurs which is the precursor of a mechanical instability 
(see also Fig.~\ref{fig:instability_resonant}).}
\label{fig:processes}
\end{figure}

Specifically, in the high-temperature regime $T  \gg  \Gamma^{\sigma}_{\alpha}$ and for high-energy separation $T \ll \varepsilon_z$, one can use an analytic approximation for the rates $\gamma^{s}_{\alpha\beta}$, which is in excellent agreement with the full results of Eq.~\eqref{eq:rates}.
The Lorentzian functions  appearing in Eq.~\eqref{eq:T_ab}  
can be treated separately as $\delta$ functions in the integral of Eq.~\eqref{eq:rates} and we can cast each rate as the sum of two spin-resolved rates 
$\gamma^{s}_{\alpha\beta} \simeq \sum_{\sigma} \gamma^{s\sigma}_{\alpha\beta}$,  
for the tunneling through the dot level with spin $\sigma=\pm$. 
The latter rates read as
\begin{eqnarray}
\label{eq:approximation}
\gamma_{\alpha\beta}^{s\sigma} &=& \frac{\lambda^2}{\Gamma_l^{\sigma}+\Gamma_r^{\sigma}}
\left\{ 
\Gamma_{\alpha}^{\sigma}\Gamma_{\beta}^{-\sigma}  T_{+}^{s\sigma}
 f_{\alpha}(\varepsilon_{\sigma})\left[1- f_{\beta}(\varepsilon_{\sigma}+s\omega)\right] 
\right. \nonumber \\
& +&  \left. 
\Gamma_{\alpha}^{-\sigma} \Gamma_{\beta}^{\sigma}  T_{-}^{s\sigma}
 f_{\alpha}(\varepsilon_{\sigma}-s\omega)\left[1- f_{\beta}(\varepsilon_{\sigma})\right]  
\right\}
\end{eqnarray}
with $T_{\pm}^{s\sigma} = 1/\left[(\Gamma_l^{-\sigma}+\Gamma_r^{-\sigma})^2+(\sigma\varepsilon_z\pm s\omega)^2\right]$. 
As we explained, for $p = 1$ the rates $\gamma^{s\sigma}_{\alpha\alpha}$ vanish, as the electron cannot return 
to its original lead after a spin-flip. Additionally, since $\Gamma_l^+=\Gamma_r^-=0$, one 
of the two terms appearing in Eq.~\eqref{eq:approximation} is zero for the spin-resolved 
rates $\gamma_{lr}^{s\sigma}$ and $\gamma_{rl}^{s\sigma}$. Assuming symmetric contacts $\Gamma_l^-=\Gamma_r^+=\Gamma$ and setting $T_{\pm}^s=\lambda^2\Gamma/[\Gamma^2+(s\omega\pm\varepsilon_z)^2]$,
the spin-resolved rates reduce to
 \begin{align}
\gamma^{s\sigma}_{lr} 
&= 
T^{s}_{-} \, 
f_l (\varepsilon_{\sigma} - s \omega \delta_{\sigma+}) 
\left[ 1 -  f_r(\varepsilon_{\sigma}+  s \omega \delta_{\sigma-})  \right] \, , \label{eq:gamma_delta_1}
\\
\gamma^{s\sigma}_{rl} 
&= 
T^{s}_{+} \, 
f_r (\varepsilon_{\sigma} - s \omega \delta_{\sigma-}) 
\left[ 1 -  f_l(\varepsilon_{\sigma}+  s \omega \delta_{\sigma+})  \right] \, . \label{eq:gamma_delta_2}
\end{align}
The processes associated to the rates $\gamma_{lr}^{s\sigma}$ in Eq.~\eqref{eq:gamma_delta_1} are shown in Figs.~\ref{fig:processes}(a)-\ref{fig:processes}(d) for $eV>0$.

The behavior of the phonon occupation in Fig.~\ref{fig:instability} can be now understood by considering the rates in Eqs.~\eqref{eq:gamma_delta_1} and  \eqref{eq:gamma_delta_2}.
For instance, in Fig.~\ref{fig:instability}, we chose the chemical potentials as $\mu_l=\varepsilon_0$ and $\mu_r=\varepsilon_0-eV$ such that for $eV<0$ mainly  the spin-up 
level contributes to transport. 
In the high-voltage approximation $|eV| \gg (T,\omega$), 
we have  $\gamma_{lr}^{s\sigma} \ll \gamma_{rl}^{s\sigma}$. 
The damping coefficient can then be approximated by the difference of two rates as  $\gamma=\gamma_{rl}^{++}-\gamma_{rl}^{-+}$.
The electrons tunnel from the right lead to the dot and finally to the left accompanied by a spin-flip. 
Further approximating the Fermi functions in the rates $\gamma_{rl}^{\sigma +}$ by $f_r\simeq 1$ and $f_l=0$, we obtain that the damping scales as $\gamma \sim T_+^+-T_+^{-}$.
In other words, the instability of the oscillator is related to the different magnitude of the transmissions. 
When the transmission for emission  processes (heating) is larger than the transmission for absorption ones (cooling), i.e., $T_+^-\gtrsim T_+^+$, 
one obtains that the damping coefficient is negative. 

Equations \eqref{eq:gamma_delta_1} and \eqref{eq:gamma_delta_2} allow us to discuss the onset of the instability.
To determine the threshold $\gamma=0$ quantitatively, we cannot use the high voltage approximation  since the instability occurs at relatively small voltages.
In the limit of $\varepsilon_z \gg \omega$, the damping reduces to $\gamma =\gamma_{rl}^{++}-\gamma_{rl}^{-+}+\gamma_{lr}^{++}-\gamma_{lr}^{-+}$. 
Then, setting $\gamma=0$, we obtain the equation for the onset of instability for vanishing intrinsic damping $(\gamma_0=0)$. 
To leading order in $T/\varepsilon_z$ the result reads as $eV=-T\,\mbox{ln}[1+(\omega+\varepsilon_z/2)/T]$ pointing out that the critical line does not depend 
on the level position $\mu-\varepsilon_0$ as shown in Fig.~\ref{fig:instability}.

\subsubsection{Resonant regime}

\begin{figure}[t!]
\begin{center}
\includegraphics[width=0.9\columnwidth]{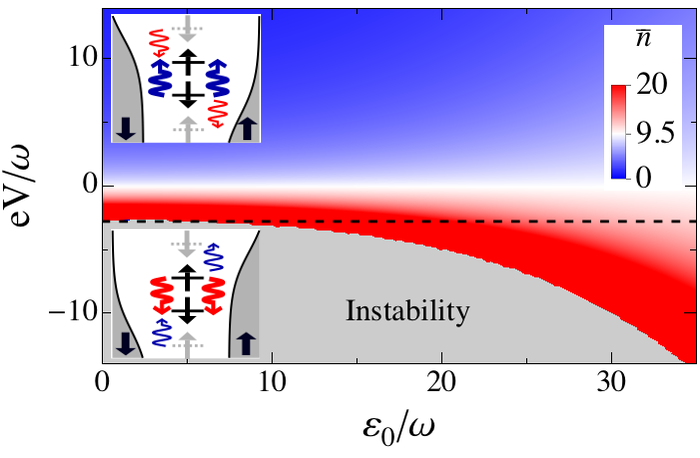}
\end{center}
\caption{(Color online) Phonon occupation $\bar{n}$ as function of  the bias voltage $eV$ and gate voltage $\varepsilon_0$. 
We consider the resonant regime $\varepsilon_z = \omega$ with
$\gamma_0=10^{-5}\omega$, $\lambda/\omega=0.01$, and $\mu_{l,r}=\varepsilon_0 \pm eV/2$. 
The other parameters are $p_l=-1$ and $p_r=1$, $\Gamma^{}_l = \Gamma^{}_r = 0.2 \omega$,
and $T= 10 \omega$. White color corresponds to $n_B(\omega)\approx 9.5$. The instability region (in gray) corresponds to $\gamma_{tot}<0$ and the black dashed line shows the analytic formula for the threshold $\gamma_{tot}=0$ (see text). The upper and lower sketches indicate the schematic behavior of the energy levels and the inelastic vibration assisted spin-flip processes which lead to cooling for $eV>0$ and heating for $eV<0$, respectively. Absorption (emission) of a vibrational energy quantum occurring in resonance is shown as blue (red) bold wiggled arrows.}
\label{fig:instability_resonant}
\end{figure}

So far, we considered a large energy splitting $\varepsilon_z\gg \omega$ without intrinsic damping. In Fig.~\ref{fig:instability_resonant}, we show the phonon occupation at resonance $\varepsilon_z=\omega$, an intrinsic quality factor
damping  $Q=10^5$, a spin-vibration coupling of $\lambda=0.01\omega$ and symmetrically applied voltage 
$\mu_{l,r} = \varepsilon_0 \pm eV/2$. 
In the resonant case, the virtual levels at energy $\varepsilon_{+} -\omega$ and $\varepsilon_{-}+\omega$ coincide, respectively, with the real dot spin levels $\varepsilon_-$ and $\varepsilon_+$ (Fig.~\ref{fig:processes} (e) and (f)). This yields a strong enhancement of the vibration assisted emission or absorption processes. 
For $eV>0$, strong cooling $\bar{n} \ll n_B(\omega)$ is achieved as discussed in Ref.~[\onlinecite{Stadler:2014hr}]. 
By reversing the voltage $eV<0$, we pass to the regime of strong heating and the oscillator becomes unstable.
This result depends on our choice for the energy of the two levels  in the dot  ($\varepsilon_{+}>\varepsilon_{-}$ for spin up and down)
and for the orientation of the left and right ferromagnets.   Changing exclusively the two levels or reversing exclusively the magnetization of the leads, 
the phonon occupation is still given by  Fig.~\ref{fig:instability_resonant}  replacing $V \rightarrow -V$.

Since now both levels are involved in transport, we have to analyze Eq.~\eqref{eq:rates} for the rates to discuss the instability. 
In the high-voltage approximation, we have again that $\gamma \simeq \delta \gamma_{lr}$ for $eV>0$ and  $\gamma \simeq \delta\gamma_{rl}$ for $eV<0$. 
In the first case, we have 
 $\delta\gamma_{lr}>0$,
such that the system remains stable.
In the second case, we found 
$\delta\gamma_{rl}<0$ 
for sufficient large voltages so that 
the damping rate $\gamma$ becomes negative.

As shown in  Fig.~\ref{fig:instability_resonant}, the system becomes unstable even for relatively low bias voltages. 
To evaluate the threshold $\gamma_{tot}=0$, we consider the high-temperature limit $T\gg (eV, \varepsilon_0, \varepsilon_z)$.
Then, we expand the Fermi functions in Eq.~\eqref{eq:rates} to lowest order in $\varepsilon/T$ and perform the integration in Eq.~\eqref{eq:rates}. 
As a result we obtain the line for which the total damping rate vanishes,
\begin{equation}
eV = - \left [ \frac{4\Gamma^2+\omega^2}{\omega} + 16\,  \gamma_0 \, T \, \Gamma \frac{4\Gamma^2+\omega^2}{\lambda^2 \omega^2} \right ]\, ,
\end{equation}
with $\Gamma_l=\Gamma_r=\Gamma$.
This line in plotted in Fig.~\ref{fig:instability_resonant} and agrees with the onset of the instability for $\varepsilon_0\lesssim T \approx 10 \omega$. 
Notice that increasing the intrinsic damping $\gamma_0$ reduces the region of instability by shifting the critical voltage to higher values.
As shown Fig.~\ref{fig:singlelead}, for larger $\varepsilon_0$, the approximation $T\gg \varepsilon_0$ gradually breaks down and becomes less accurate. In this regime, the instability line strongly depends on the intrinsic damping.

\subsection{Single polarized lead}
\label{subsec:singlepol}
In the previous section, we discussed the phonon occupation for fully polarized ferromagnetic leads. 
A finite polarization reduces the vibration-assisted spin-flip rates $\gamma_{lr}^s$ and $\gamma_{rl}^s$ in comparison 
to those rates at fully polarized ferromagnets.
Additionally, as shown in Eq.~\eqref{eq:n_bar2}, we have to consider the vibration-assisted spin-flip processes involving a single lead, 
with rates $\gamma_{ll}^s$ and $\gamma_{rr}^s$, which drive the oscillator to thermal equilibrium.

\begin{figure}[t!]
\begin{center}
\includegraphics[width=0.9\columnwidth]{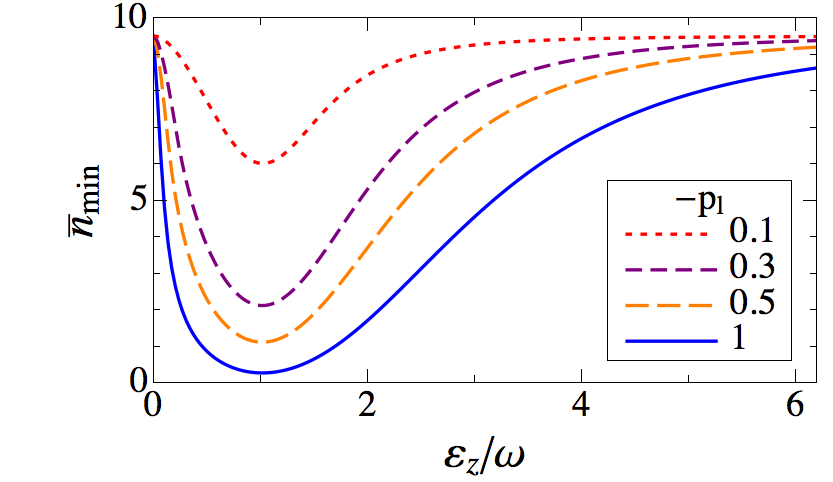}
\end{center}
\caption{(Color online) Minimal phonon occupation on the surface $(\varepsilon_0,eV)$ for a polarized  left lead with $-1\le p_l <0$ and a normal right lead as a function of $\varepsilon_z/\omega$. The parameters are  $\Gamma_l=\Gamma_r=0.2\omega$, $T=10\omega$, $Q=10^4$ and $\lambda=0.05\omega$. The minimum of $\bar{n}_{min}$ is approached at resonance $\varepsilon_z=\omega$ with $\bar{n}_{min} \simeq 0.2$ for a fully polarized left lead $p_l=-1$. }
\label{fig:singlelead}
\end{figure}

In this section, we show that active cooling can be achieved even for a single polarized lead. We assume a polarized left lead ($-1\le p_l<0$) and a normal right lead ($p_r=0$).
In Fig.~\ref{fig:singlelead}, we show the result for the minimum of the phonon occupation $\bar{n}_{min}$ on the surface ($\varepsilon_0,eV$) as a function of the energy separation  $\varepsilon_z$. We remark that, for a single polarized lead, ground-state cooling is achieved with $\bar{n}_{min} \ll 1$ at resonance $\varepsilon_z=\omega$.
The reason for the strong cooling can be understood by considering the phonon occupation Eq.~\eqref{eq:n_bar2} which in the high-voltage approximation $eV \gg (T,\omega)$ can be written as
\begin{equation}
\label{eq:n_finite_pol}
\bar{n} \simeq  
\frac{
(\gamma_0+\delta\gamma_{ll}^{\phantom{g}}+\delta\gamma_{rr}^{\phantom{g}})n_B(\omega)+\gamma^{-}_{lr}
} 
{
\gamma_0 +  \delta\gamma_{ll}^{\phantom{g}} + \delta\gamma_{rr}^{\phantom{g}} +\delta\gamma_{lr}^{\phantom{g}} 
}.
\end{equation}
At $p_l=-1$ the rates $\delta\gamma_{ll}$ are zero and only the spin-flip processes at the right lead with a rate $\delta\gamma_{rr}$ are active to drive the oscillator towards thermal equilibrium.
However, such processes give a relevant contribution to the damping $\gamma$ only if one of the two spin levels is aligned or close to the right chemical potential $\mu_r$.  
Therefore, if we have, for instance, $\varepsilon_{\pm} \gg \mu_r$, these processes are strongly suppressed and ground-state cooling can still be achieved at resonance.
In a simple picture, the left lead acts essentially as a source of spin-polarized electrons with the same spin orientation of the lower energy-level in the dot.
However, at finite polarization of the left lead, $-1<p_l<0$,  the spin-flip processes at the left lead are unavoidable $(\delta\gamma_{ll}^{} \neq 0)$  so that cooling is reduced. 

Concerning the state of the oscillator, the configuration $-1\le p_l<0$ and $p_r= 0$ discussed so far is equivalent to the configuration $p_l=0$ and $0<p_r\le 1$. 
Since the left lead is a normal metal, both the spin-up and spin-down level can be occupied by an electron tunneling from the left to the dot's levels at voltages $eV>0$. Assuming for simplicity the resonant case $\varepsilon_z=\omega$ and $p_r=1$, the right lead then selects only spin-up electrons.  The process of an absorption of a vibrational energy quantum, where the spin-down electrons flips the spin and tunnels to the right lead, occurs in resonance and thus leads again to active cooling of the oscillator.

\subsection{Parallel magnetization configuration}
\label{subsec:parallelconfiguration}

We briefly summarize the result obtained in the parallel magnetization configuration. For fully polarized leads $p_{r}=p_{l}=\pm 1$, all the vibration-assisted inelastic spin flip rates in Eq.~\eqref{eq:rates} vanish since these rates are proportional to the products $\Gamma_{\alpha}^+\Gamma_{\beta}^-=0$. An electron can neither tunnel from one lead to the other lead nor to the initial lead accompanied by an inelastic spin flip and, according to Eq.~\eqref{eq:n_bar}, the oscillator remains at equilibrium $\bar{n}=n_B(\omega)$.
At finite but equal polarization $ p_{r} = p_{l}  \neq \pm1$, it is instructive to compare the majority and minority charge carriers involved in an inelastic tunneling event in the parallel and antiparallel configuration. 
In both parallel and antiparallel configurations, the processes associated to
the rates $\gamma_{\alpha\alpha}^s$ connect the majority spin carriers with the minority carriers of the same lead. 
The processes associated to the rates $\gamma_{lr}^s$ and $\gamma_{rl}^s$ in the antiparallel 
configuration connect the majority carriers from one lead with the majority carriers of the opposite lead. 
However, in the parallel configuration, the rates $\gamma_{lr}^s$ and $\gamma_{rl}^s$ connect the majority spin carriers with the minority carriers
leading to a suppression of these rates. In the parallel configuration, we found that an applied voltage increases the phonon occupation $\bar{n}>n_B(\omega)$ and active cooling does not occur.

\section{Current}
\label{sec:current}
In this section, we investigate the influence of the spin-vibration interaction and the resulting nonequilibrium phonon occupation on the current through the quantum dot.
To this end, we calculate the corrections to the current to the leading order in the spin-vibration coupling and for two different cases. 
 
In Sec.~\ref{sec:current_expansion}, we explain the general expansion for the current.
In Sec.~\ref{sec:current_equilibrated}, we assume that the oscillator is strongly coupled to the external bath such that 
$\gamma_0 \gg \gamma$. 
Then, the time for thermal relaxation is much smaller than the time associated to the inelastic spin-flip processes to set  the oscillator in an unequilibrated state. 
The oscillator is mainly in an equilibrated state and its phonon occupation can be described by the Bose distribution function. 
This regime is referred as the regime of thermal or equilibrated vibration. 
In Sec.~\ref{sec:current_unequilibrated}, we consider the regime $\gamma_0 \ll \gamma$.
Then, as we discussed in the previous section, the oscillator is driven by the current itself towards a nonequilibrium phonon occupation.
This regime is referred as the regime of nonequilibrated vibration.

\subsection{General expansion for the current}
\label{sec:current_expansion}

The current is obtained by inserting the Keldysh and retarded elements of the expansion \eqref{eq:checkG} in the expression for the current \eqref{eq:I_current_2}.
The result can be written as an elastic current $I_0$ in the absence of the spin-vibration coupling, 
an elastic and an inelastic correction $I_{ec}$ and $I_{in}$,
\begin{equation}
\label{eq:current}
I = I_0+I^{}_{ec}\left[ \check{\Sigma} \right]+I^{}_{in}\left[ \check{\Sigma} \right]  .
\end{equation}
The result for the elastic current corresponds to the well-known formula
\begin{equation}
I_{0} \mathord= \frac{e}{h}\int d\varepsilon \sum_\sigma 4\Gamma_l^\sigma \Gamma_r^\sigma \vert G^R_\sigma(\varepsilon) \vert^2 \left(f_l(\varepsilon)\mathord-f_r(\varepsilon)\right),
\label{eq:elasticcurrent} 
\end{equation}
with the Green's function given by the Eqs. \eqref{eq:dotgreensfunction_1} and \eqref{eq:dotgreensfunction_2}.
Both $I^{}_{ec}$ and $I^{}_{in}$  are proportional to $\lambda^2$ and they are functional
of the electron self energies $\check{\Sigma}$ appearing in Eqs.~\eqref{eq:Sigma_RA} and \eqref{eq:Sigma_K}
which are related to the phonon Green's function $\check{D}$.

\subsection{Current with equilibrated vibration}
\label{sec:current_equilibrated}
\setcounter{paragraph}{0}
Assuming $\gamma_0 \gg \gamma$, we approximate $\gamma_{tot} \simeq \gamma_0$ in Eq.~(\ref{eq:D_RA}) for the retarded/advanced component 
of the phonon Green's function.  
Correspondingly, we neglect 
$\Pi^K$ (related to the spin-vibration interaction) as $|\Pi^K| \ll |\Sigma_0^K|$  in Eq.~(\ref{eq:D_K}) for the Keldysh component of the phonon Green's function.
Inserting the resulting phonon Green's function in the electronic self-energies in Eqs.~\eqref{eq:Sigma_RA} and \eqref{eq:Sigma_K}, we calculate the currents 
$I^{}_{ec}$ and $I^{}_{in}$ in Eq.~(\ref{eq:current}).
In the remaining part of this section, we discuss separately the elastic corrections to the linear conductance
and the inelastic corrections to the differential conductance. 
Thereby, we mainly focus on the characteristic features 
of the ferromagnetic leads and the spin-vibration interaction in the transport.

\subsubsection{Elastic correction with equilibrated vibration}
\begin{figure}[btp]
\begin{center}
\includegraphics[width=0.9\columnwidth]{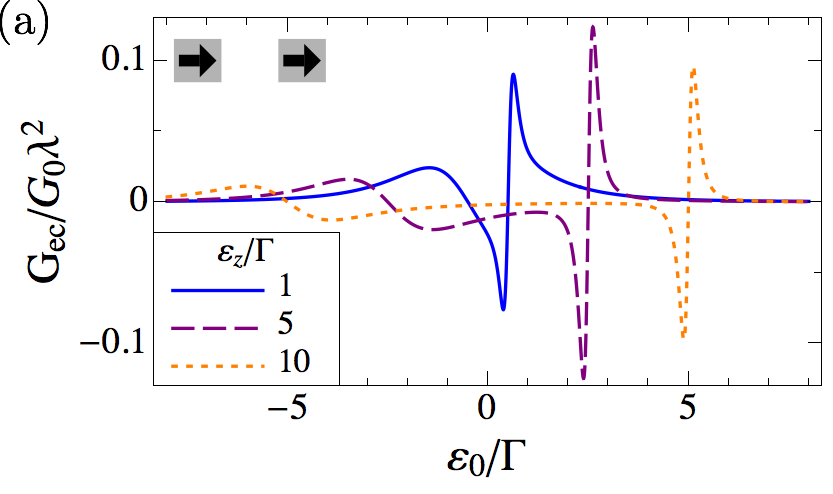}
\includegraphics[width=0.9\columnwidth]{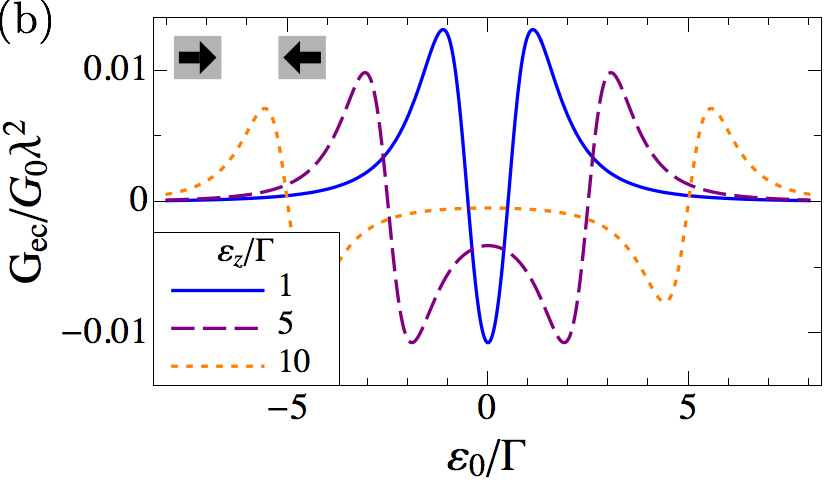} 
\caption{(Color online) Elastic correction to the linear conductance at $T=0$, symmetric coupling $\Gamma_l=\Gamma_r=\Gamma$, $\omega=5\Gamma$, and zero chemical potentials. (a) Parallel magnetization configuration ($p_r=p_l=0.8$). (b) Antiparallel configuration ($p_r=-p_l=0.8$). The different magnetization configurations in (a) and (b) lead to $G_{ec}(\varepsilon_0)\neq G_{ec}(-\varepsilon_0)$ in (a) whereas the correction is symmetric $G_{ec}(\varepsilon_0)=G_{ec}(-\varepsilon_0)$  in (b).  In the range $\varepsilon_+\varepsilon_- <0$ the correction to the conductance is negative (see text). 
}
\label{fig:elastic_correction}
\end{center}
\end{figure}

The elastic correction of our model Hamiltonian can be written as  ($\varepsilon_s = \varepsilon+s\omega$)
\begin{align}
I_{ec} = \frac{e}{h} \int d\varepsilon \sum_\sigma 8 \Gamma_l^\sigma \Gamma_r^\sigma \vert G^R_\sigma(\varepsilon) \vert^2 & \nonumber \\  \mbox{Re}[G_\sigma^R(\varepsilon) \Sigma_{-\sigma-\sigma}^R(\varepsilon)]   &[f_l(\varepsilon)-f_r(\varepsilon)] \, .
\label{eq:elasticcorrection}
\end{align}
We focus the discussion on the linear conductance at $T=0$. 
In this case, the retarded self-energy $\Sigma^R_{\sigma\sigma}$ inside the integral of Eq.~\eqref{eq:elasticcorrection} 
can be calculated analytically 
in the limit $\gamma_0 \ll (\omega,\Gamma_l,\Gamma_r,eV)$
and the explicit expression is given in Appendix \ref{Appendix:retardedselfenergy}. The correction to the linear conductance $G=dI_{ec}/dV \vert_{V=0}$ reduces to 
\begin{multline}
\frac{G_{ec}}{G_0\lambda^2} \mathord= \sum_{s\sigma} \frac{ 2  \Gamma^{-\sigma}\Gamma_l^\sigma \Gamma _r^{\sigma}\varepsilon_\sigma }{{({\Gamma^{\sigma}}^2\mathord+\varepsilon_\sigma^2)}^2 {[\Gamma^{-\sigma}}^2\mathord+\left(\varepsilon_{-\sigma}\mathord+s\omega\right){}^2]} \left(\frac{\varepsilon_{-\sigma}\mathord+s\omega}{\Gamma^{-\sigma}} \right. \\ \left. \left( 1\mathord+\frac{2 s }{\pi} \mathrm{tan}^{-1}\left(\frac{\varepsilon_{-\sigma}}{\Gamma^{-\sigma}}\right)\right) \mathord-\frac{s}{\pi}\mathrm{ln}\left(\frac{\omega^2}{\varepsilon_{-\sigma}^2+{\Gamma^{-\sigma}}^2}\right)\right),
\label{eq:elasticconductance}
\end{multline}
with $G_0=2e^2/h$, $\Gamma^\sigma = \Gamma_l^\sigma+\Gamma_r^\sigma$ and $\mu_l=\mu_r=0$.
In Figs.~\ref{fig:elastic_correction}(a) and (b), we show the correction to the elastic conductance for the parallel ($p_r=p_l=0.8$) and 
antiparallel ($p_r=-p_l=0.8$) configuration with $\Gamma_l=\Gamma_r=\Gamma$. 

In the parallel configuration, we observe that the correction at $G_{ec}(\varepsilon_0)$ differs from the corrections at $G_{ec}(-\varepsilon_0)$ whereas we find that $G_{ec}(\varepsilon_0)=G_{ec}(-\varepsilon_0)$ in the parallel configuration. 
Such a behavior is explained by the polarization of the ferromagnetic leads.
In the parallel configuration, the spin-up level is coupled stronger to the leads than the spin-down level. The different couplings lead to sharp features in the correction to the conductance close to the spin-down level, whereas close to the spin-up level the correction is broadened. 
On the contrary, for the antiparallel configuration, there are always electrons of the majority and minority spin involved when an electron tunnels from the left to the right lead. This gives rise to the symmetric 
behavior $G_{ec}(\varepsilon_0)=G_{ec}(-\varepsilon_0)$.

We notice that the elastic correction to the conductance in Fig.~\ref{fig:elastic_correction} can be either positive or negative as varying $\varepsilon_0$ both for the parallel and antiparallel configurations. 
Such a behavior is different from the results obtained in the Anderson-Holstein model for 
a spinless dot level in which the sign of the conductance corrections does not change to the first leading order in the electron-vibration coupling.\cite{Egger:2008hh}   
The negative correction to the conductance occurs due to Fano interference effects. At finite polarization, an electron with spin $\sigma$ can pass the quantum dot through two different paths. The first path corresponds the elastic tunneling of an electron with spin-up (-down) through the spin-up (-down) level without interacting with the oscillator. The second path is associated to the spin-vibration interaction. For instance, an electron of spin $\sigma$ can also tunnel elastically from one lead to the other lead by flipping its spin and virtually exciting the oscillator. The latter is excited by an emission (absorption) of a vibrational energy quantum followed by an absorption (emission) of a vibrational energy quantum so that the electron ends up at the same energy of its initial state [see Fig.~\ref{fig:self-energy}(b)]. In the range $\varepsilon_+ \varepsilon_-<0$, the spin-up level is above the Fermi energy and particle-like processes contribute to the correction whereas the spin-down level is below the Fermi energy and hole-like processes dominate. The transmission amplitude of the electronlike and holelike paths differ by a phase of $\pi$ 
leading to the negative correction to the conductance in the range  $\varepsilon_+ \varepsilon_-<0$.\cite{Kubala:2003bo,Kubala:2002bh}

\subsubsection{Inelastic current with equilibrated vibration}

In the limit $\gamma_0 \ll (\omega,\Gamma_l,\Gamma_r,T,eV)$ , the inelastic current can be written in terms of the rates $\gamma_{\alpha\beta}^s$ of Eq.~\eqref{eq:rates} as
\begin{equation}
\label{eq:inelasticcurrent}
I_{in}=\frac{2e}{\hbar} 
\left[ 
\left(n_B(\omega)+1\right)  \left(\gamma_{lr}^{-}-\gamma_{rl}^{-}\right)
+
n_B(\omega) \left(\gamma_{lr}^{+}-\gamma_{rl}^{+}\right)
\right].
\end{equation}
Transport is possible via the emission and the absorption of vibrational energy quanta. At zero temperature, $n_B(\omega)=0$, and the threshold voltage for having  an emission of a vibrational energy quantum is $eV = \omega$. 
Note that, as we calculated the inelastic current to the leading order in the coupling, only single-phonon 
processes are taken into account in Eq.~(\ref{eq:inelasticcurrent}). The differential conductance 
$G_{in}=dI_{in}/dV$ at zero temperature can be written as
\begin{equation}
\frac{G_{in}}{G_0\lambda^2}=  \sum_{\sigma \alpha} \Gamma_{\alpha}^{\sigma}\Gamma_{-\alpha}^{-\sigma}\vert G^R_{\sigma}(\mu_{\alpha}) G^R_{-\sigma}(\mu_{\alpha}-\alpha\omega)\vert^2\theta(\mu_l-\mu_r-\omega),
\end{equation}
with $(\alpha,\beta)=(l,r)=\pm$ and the retarded Green's function given by Eq.~(\ref{eq:dotgreensfunction_1}).
\begin{figure}[!t]
\begin{center}
\includegraphics[width=0.9\columnwidth]{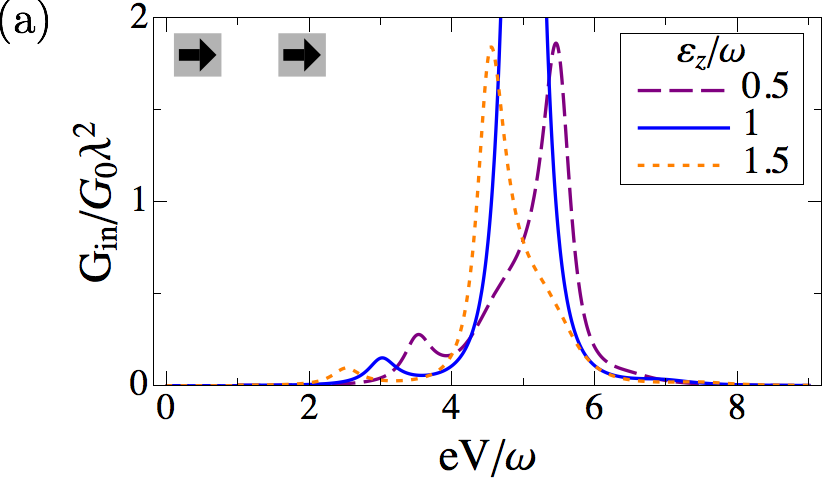}
\includegraphics[width=0.9\columnwidth]{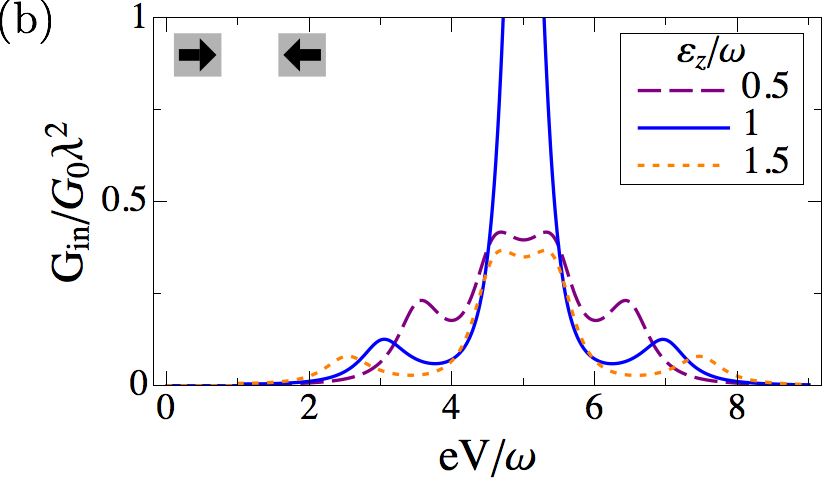} 
\end{center}
\caption{(Color online) Inelastic contribution to differential conductance at 
zero temperature and $\varepsilon_0=2\omega$, $\Gamma_l=\Gamma_r=0.2\omega$, and symmetrically applied voltage. In (a), the polarization of the ferromagnetic leads are aligned parallel $p_l=p_r=0.4$. In (b), we show the antiparallel configuration with $p_r=-p_l=0.4$. The peaks of the inelastic differential conductance appear at voltages $eV/2=\varepsilon_\pm$ and $eV/2=\varepsilon_\pm+\omega$ (see text).
 }
\label{fig:inelasticcurrent}
\end{figure}
Figures \ref{fig:inelasticcurrent} (a) and (b) show the inelastic differential conductance at zero temperature in the parallel and antiparallel configurations, respectively. 
The voltage is applied symmetrically $\mu_l=eV/2$ and $\mu_r=-eV/2$, the energy level on the dot is set to $\varepsilon_0=2\omega$, and the polarization is $p=p_r= p_l  =0.4$ for the parallel configuration and $p=p_r= -p_l  =0.4$ for the antiparallel configuration.
In Fig.~\ref{fig:inelasticcurrent}, the inelastic processes can occur at the voltages $eV/2=\varepsilon_\pm $ and $eV/2=\varepsilon_\pm+\omega$.
To illustrate the behavior of these inelastic peaks, we discuss in details the antiparallel case shown in Fig.~\ref{fig:inelasticcurrent}(b) for 
$\varepsilon_z=1.5\omega$.
The first peak appears due to the resonance of the left Fermi level with the spin-down level on the quantum dot ($eV/2=\varepsilon_{-}$).
In this case, a spin-down electron is transferred to the quantum dot followed by a spin flip and an emission of a vibrational energy quantum when it moves to 
the right barrier [see the schematic picture in Fig.~\ref{fig:processes}(d)].
At higher voltage, a second peak appears at $eV/2=\varepsilon_{-}+\omega$. 
In this case, a spin-up electron tunneling from the left lead can enter the dot spin-down level by emitting a vibrational energy quantum.
Similar processes occur at higher voltage when the Fermi energy in the left lead is in resonance with the spin-up level 
of the quantum dot  $eV/2=\varepsilon_{+}$ or at the voltage $eV/2=\varepsilon_{+}+\omega$, the latter case reported 
in Fig.~\ref{fig:processes}(b).
At resonance $\varepsilon_z=\omega$, two peaks merge into a single peak and the differential conductance is strongly increased compared to the case out of resonance. 
Following similar arguments reported in Sec.~\ref{subsec:instability}, this is due to the virtual level $\varepsilon_{-}+\omega$ coinciding with the real dot level $\varepsilon_{+}$.

\subsection{Current with unequilibrated vibration}
\label{sec:current_unequilibrated}

As a next step, we discuss the current for the case of unequilibrated vibration for antiparallel ferromagnetic leads.
\begin{figure}[t!]
	\begin{center}
		\includegraphics[width=0.46\columnwidth]{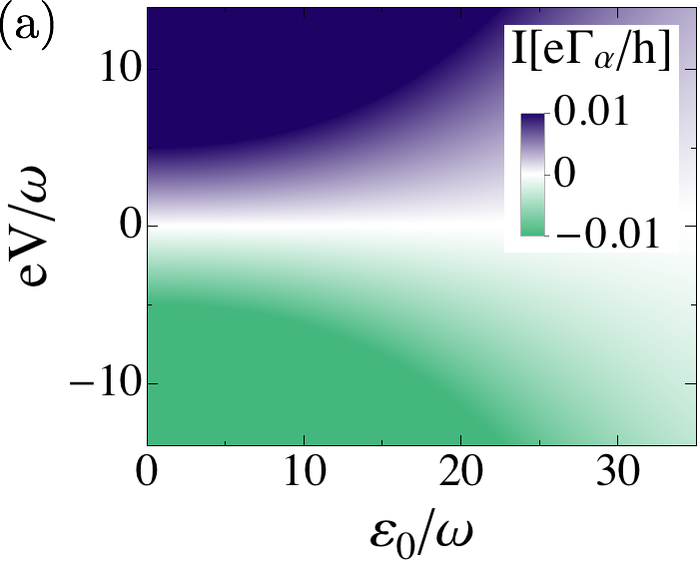} \quad
		\includegraphics[width=0.46\columnwidth]{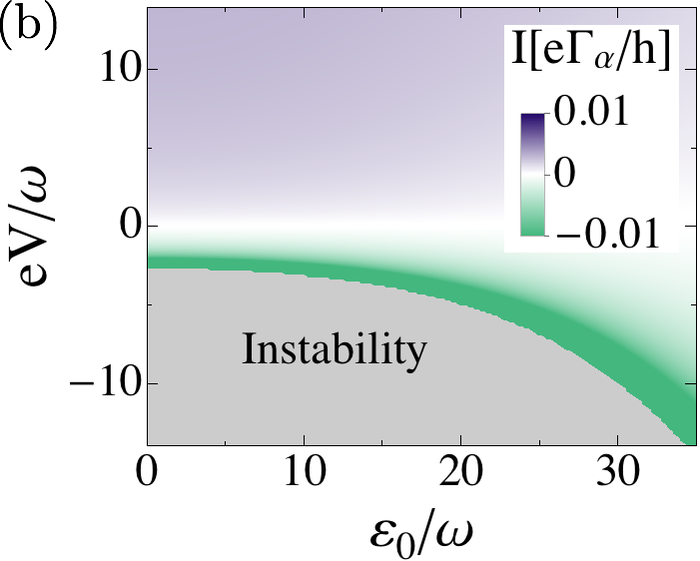} 
	\end{center}
	\caption{ (Colors online) Inelastic current for fully polarized antiparallel ferromagnets ($p_r=-p_l=1$) at resonance $\varepsilon_z=\omega$, $T=10\omega$, and $\Gamma=0.2\omega$. (a) Equilibrated vibration with $\bar{n}=n_B(\omega)$. (b) Unequilibrated vibration with a coupling constant $\lambda=0.01\omega$ and an intrinsic damping of $\gamma_0=10^{-5}\omega$. The nonequilibrium phonon occupation $\bar{n}$ corresponding to the inelastic current in (b) is shown in Fig.~\ref{fig:instability_resonant}. For $eV>0$, the oscillator is strongly cooled $\bar{n}\ll n_B(\omega)$, leading to a suppression of the inelastic current in (b) compared to the case of equilibrated vibration in (a). 
	}
	\label{fig:nonequiinelasticcurrent}
\end{figure}
We found clear signatures of the nonequilibrium phonon occupation in terms of a suppression (enhancement) 
of the current when the phonon occupation of the oscillator decreases (increases)  compared to thermal vibration.

For the regime of unequilibrated vibration, we use the full phonon Green's functions \eqref{eq:D_RA}  and \eqref{eq:D_K} to calculate 
the electron self-energies (\ref{eq:Sigma_RA}) and (\ref{eq:Sigma_K}) and, hence, the current
Eq.~(\ref{eq:current}) in the limit $\gamma_{tot} \ll (\omega,\Gamma_l,\Gamma_r,T,eV)$.
The results are similar to the previous case for the equilibrated vibration. For instance, the inelastic correction $I_{in}$
is similar to \eqref{eq:inelasticcurrent} in which we have to replace the thermal phonon occupation $n_B(\omega)$ 
with the nonequilibrium occupation $\bar{n}$ as given by Eq.~(\ref{eq:n_bar}).
For oscillators with very high quality factor, we have that $\bar{n}$ is essentially $n$, as given by Eq.~(\ref{eq:n_bar}).
A similar approach was used in other nanomechanical systems.\cite{Walter:2011kw}
We consider such an approach reasonable for weak spin-vibration coupling and low current through the dot.
At the same time, it is also useful to discuss qualitatively the behavior of the system to understand the possible features 
appearing in the current-voltage characteristic associated to a strongly cooled or heated oscillator.
A more refined self-consistent approach, as discussed in Ref.~[\onlinecite{Skoldberg:2008ge}],  is beyond the aim of this work.

\begin{figure}[b!]
	\begin{center}
		\includegraphics[width=0.46\columnwidth]{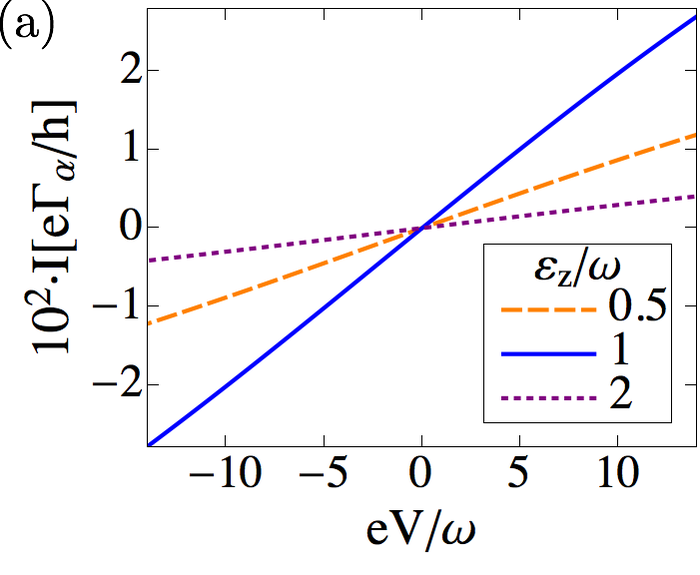} \quad
		\includegraphics[width=0.46\columnwidth]{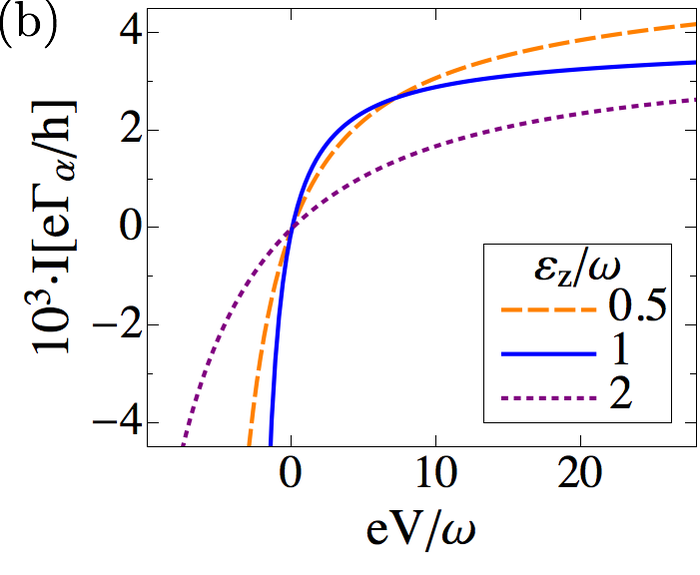} \quad
		\includegraphics[width=0.46\columnwidth]{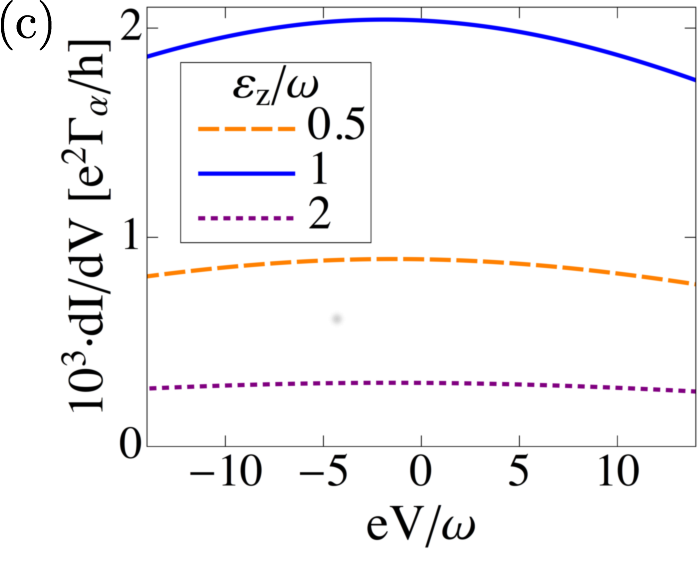} \quad
		\includegraphics[width=0.46\columnwidth]{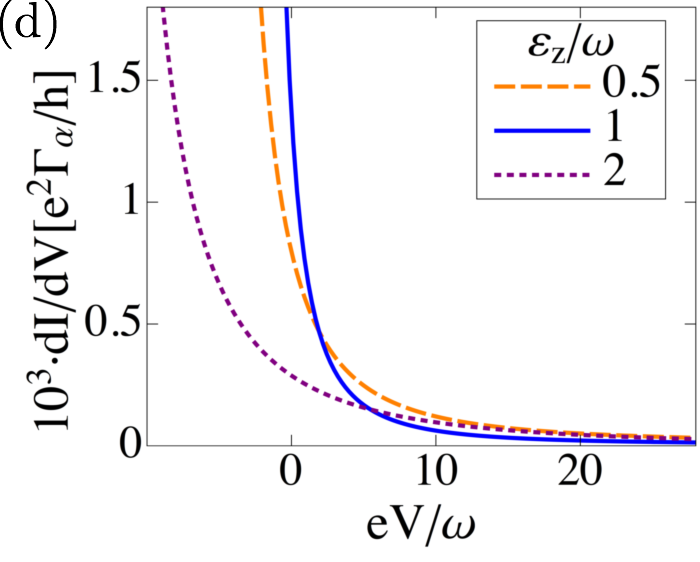}
	\end{center}
	\caption{(Color online) Current for equilibrated (a) and unequilibrated vibration (b) for fully antiparallel polarized ferromagnets $p_r=-p_l=1$, $T=10\omega$, $\varepsilon_0=0$, 
		$\Gamma=0.2\omega$, $\lambda=0.01\omega$, and $\gamma_0=10^{-5}\omega$. For $eV>0$, the current in (b) is  suppressed compared to the current in (a). At negative voltages in (b), the oscillator approaches the mechanical instability and sharply decreases. In (c) and (d), we show the differential conductance $dI_{in}/dV$ corresponding to (a) and (b), respectively.}
	\label{fig:nonequiinelasticcurrent1dfully}
\end{figure}

We start with the discussion of fully polarized leads in the antiparallel configuration $p_r=-p_l=1$. Notice that, in this case, the elastic contributions to the current vanish: $I_0 =0$ [Eq.~\eqref{eq:elasticcurrent}] and 
$I_{ec} =0$ [Eq.~\eqref{eq:elasticcorrection}], since the electrons have to change their spin when tunneling from one lead to another.
This can happen only through inelastic phonon-assisted  spin-flip processes.
Therefore, the total current  Eq.~\eqref{eq:current} reduces to the inelastic current $I_{in}$ given by Eq.~\eqref{eq:inelasticcurrent} 
with $n_B(\omega)$ replaced by $\bar{n}$.
At resonance $\varepsilon_z=\omega$,  such inelastic processes can cool the oscillator, $\bar{n} \ll n_B(\omega)$, for positive voltage $eV>0$ [see Fig.~\ref{fig:processes}(e)], whereas they can heat the oscillator, $\bar{n} \gg n_B(\omega)$, for negative voltage $eV<0$ [see Fig.~\ref{fig:processes}(f)].

\begin{figure}[t!]
	\begin{center}
		\includegraphics[width=0.46\columnwidth]{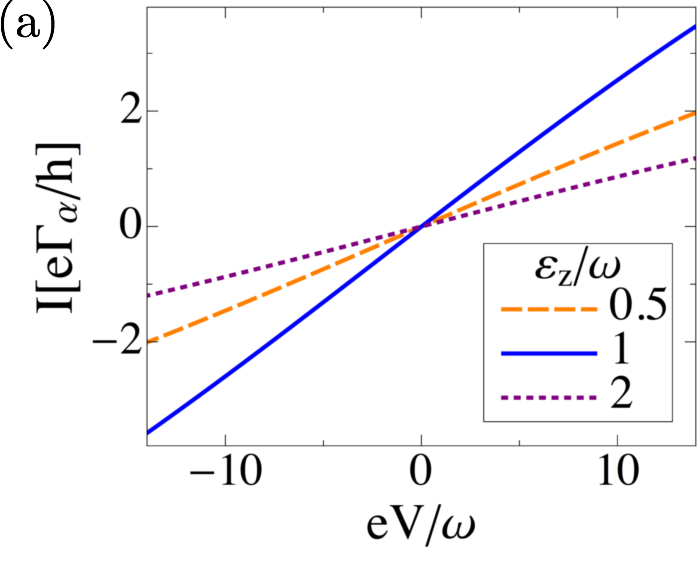} \quad
		\includegraphics[width=0.46\columnwidth]{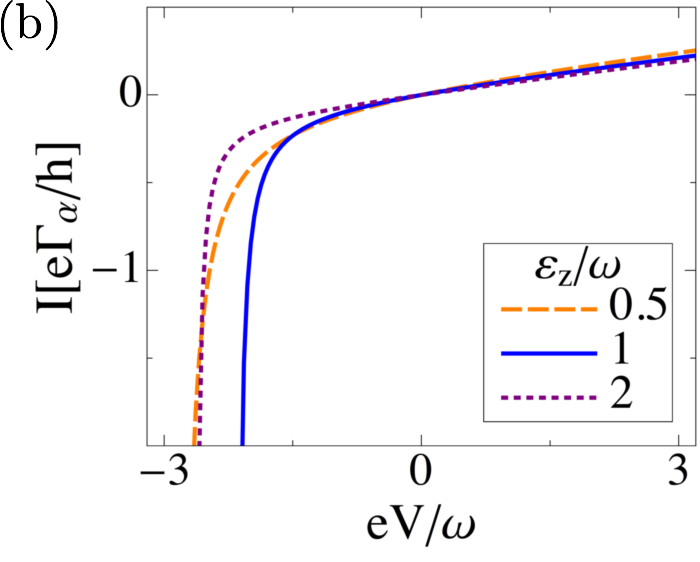} \quad
		\includegraphics[width=0.46\columnwidth]{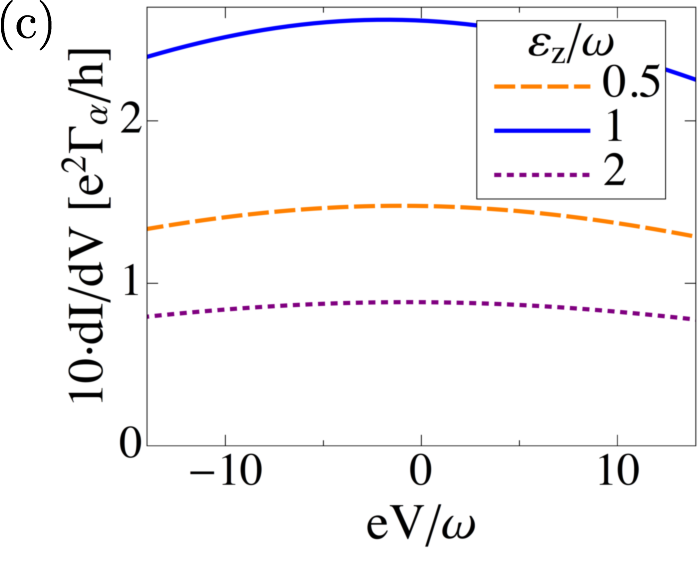} \quad
		\includegraphics[width=0.46\columnwidth]{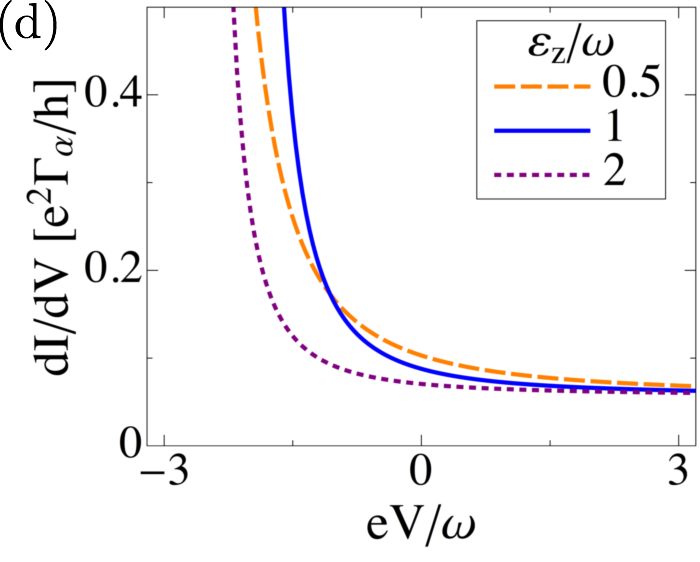}
	\end{center}
	\caption{(Color online) Current and differential conductance for equilibrated [(a) and (c)] and unequilibrated vibration [(b) and (d)] at polarization $p_r=-p_l=0.5$, $T=10\omega$, $\varepsilon_0=0$, $\Gamma=0.2\omega$, $\lambda=0.2\omega$, and $\gamma_0=10^{-5}\omega$. In (b) the current at $eV<0$ decreases since the oscillator approaches the mechanical instability and the phonon occupation strongly increases. At $eV>0$, the current is suppressed compared to the current in (a) since $\bar{n}<n_B(\omega)$. In (c) and (d), we show the differential conductance $dI/dV$ corresponding to (a) and (b), respectively. }
	\label{fig:nonequiinelasticcurrent1d}
\end{figure}

In Fig.~\ref{fig:nonequiinelasticcurrent}(a) and (b), we compare the current at resonance for equilibrated $[\gamma_0 \gg \gamma, \bar{n}\simeq n_B(\omega)]$
and unequilibrated vibration $[\gamma_0 \ll \gamma,  \bar{n}\simeq n]$ for fully polarized antiparallel ferromagnets at finite temperature $T=10\omega$. 
Essentially, the nonequilibrium phonon occupation corresponding to the current of Fig.~\ref{fig:nonequiinelasticcurrent}(b) is shown in Fig.~\ref{fig:instability_resonant}.  
For $eV>0$, the current in Fig.~\ref{fig:nonequiinelasticcurrent}(b) is strongly suppressed compared to the case of equilibrated vibration.
In this case, the oscillator is cooled close to its quantum ground state $(\bar{n}\ll 1)$ so that electrons can tunnel only through phonon-emission 
characterized by the rate $\gamma_{lr}^{-}$ [see Eq.~\eqref{eq:inelasticcurrent}].
Since in the cooling regime the relation $\gamma_{lr}^{-} \ll \gamma_{lr}^{+}$ holds, the current flowing through the dot results extremely low.
In other words, increasing the current implies cooling the oscillator more efficiently which turns out in a reduction of the current itself.
By contrast,
for $eV<0$, the current strongly decreases with the voltage for unequilibrated vibration before the regime of instability is reached at some threshold voltage.

In Figs.~\ref{fig:nonequiinelasticcurrent1dfully} (a) and (b), we show the inelastic current for equilibrated and unequilibrated vibration and different energy separation $\varepsilon_z$. The parameters are the same as in Fig.~\ref{fig:nonequiinelasticcurrent} but the current is shown as a function of voltage at $\varepsilon_0=0$. 
In Fig.~\ref{fig:nonequiinelasticcurrent1dfully}(a), the vibration-assisted spin-flip rates give the largest contribution to the current at resonance $\varepsilon_z=\omega$. In Fig.~\ref{fig:nonequiinelasticcurrent1dfully} (b), we find that the current for $eV>0$ is strongly suppressed compared to the current in Fig.~\ref{fig:nonequiinelasticcurrent1dfully}(a). For $eV<0$, the current sharply decreases since the oscillator approaches the mechanical instability. Out of resonance, the current 
decreases at larger negative voltages compared to the resonant case.
The differences between equilibrated and unequilibrated vibration are also visible in the differential conductance $dI_{in}/dV$ shown in  Figs.~\ref{fig:nonequiinelasticcurrent1dfully} (c) and (d).

As last point, we analyzed the effects of a finite polarization in the current-voltage characteristic. 
We calculated the full current as given by Eq.~\eqref{eq:current} including the leading elastic term $I_0$ [Eq.~\eqref{eq:elasticcurrent}]
and the elastic correction $I_{ec}$ [Eq.~\eqref{eq:elasticcorrection}].
In Figs.~\ref{fig:nonequiinelasticcurrent1d} (a) and (b), we compare the current for equilibrated and unequilibrated vibrations at $p_r=-p_l=0.5$. 
Here, we set $\lambda=0.2\omega$ ($\omega=2\pi \cdot 100 \textrm{ MHz}$) for large spin-orbit coupling estimated by recently reported measurement 
for the spin-orbit coupling $\Delta_{SO}$ in carbon nanotubes.\cite{Steele:2013eq} 
We can still observe the strong suppression of the current at $eV>0$ compared to equilibrated vibration in Fig.~\ref{fig:nonequiinelasticcurrent1d} (a) as well as the sharp decrease of the current when the oscillator approaches the mechanical instability. 
At positive voltages in Fig.~\ref{fig:nonequiinelasticcurrent1d}(b), the current is dominated by the elastic current $I_0$, since the oscillator is strongly cooled. The sharp decrease at negative voltages occurs due to the corrections to the current. Note the different scales of the current in Fig.~\ref{fig:nonequiinelasticcurrent1d} and \ref{fig:nonequiinelasticcurrent1dfully}. The differential conductances corresponding to the current in  Fig.~\ref{fig:nonequiinelasticcurrent1d} (a) and (b), are shown in Fig.~\ref{fig:nonequiinelasticcurrent1d} (c) and (d), respectively.

Summarizing this section we note that the current follows the nonequilibrium phonon occupation in some regime.
Since the current strongly depends on the polarization and alignment of the ferromagnets, transport measurements with tunable nano-ferromagnetic contacts can provide a feasible way to detect the spin-vibration interaction in suspended CNTQDs.

\section{Summary}
\label{sec:summary}
For a suspended CNTQD in a spin-valve geometry, we studied the spin-dependent current 
through two spin levels and the steady-state phonon occupation for a vibrational flexural mode 
in presence of a spin-vibration interaction.
Such a spin-vibration interaction is caused by the spin-orbit coupling or a magnetic gradient.
We have shown that even weakly spin-polarized currents allow the control of the phonon occupation $\bar{n}$ in a way that a flexural mode can be cooled $[\bar{n} \ll n_B(\omega)]$ or heated $[\bar{n} \gg n_B(\omega)]$ 
or even driven towards a mechanical instability regime in which the mechanical damping becomes negative.
Such a control can be achieved by manipulating several parameters of the system.
In particular, it can be obtained using electrical fields, viz. varying the bias-voltage polarity or the gate voltage, 
or using magnetic fields, viz. by changing the orientation of the magnetic polarization of the ferromagnetic contacts 
or tuning the energy separation of the dot's spin levels.
The current shows characteristic features of the nonequilibrium phonon occupation and directly can be exploited to
demonstrate the presence of the spin-vibration interaction and the non-thermal phonon occupation of the oscillator.

\acknowledgments 
This research was kindly supported by the EU FP7 Marie Curie Zukunftskolleg Incoming Fellowship Programme, University of Konstanz (Grant No. 291784) and by the DFG through the Collaborative Research Center SFB 767 and through the project BE 3803/5. 
 
\appendix

\section{Phonon self-energy for the vibration-environment coupling}
\label{Appendix:Phononenvironment}

We consider a mechanical oscillator coupled to the environment which is described as 
an ensemble of independent harmonic oscillators (the Caldeira-Leggett model).
The Hamiltonian of the external environment reads 
\begin{equation}
\hat{H}_{env} =   (\hat{b}^{\dagger} +  \hat{b}^{\phantom{g}}) 
\sum_k \lambda_k (\hat{b}^{\dagger}_k +  \hat{b}_k^{\phantom{g}})
+
\sum_k \omega_k  \hat{b}^{\dagger}_k \hat{b}_k^{\phantom{g}}\,.
\end{equation}
As the Hamiltonian is bilinear, the model is exactly solvable: The phonon self-energy $\check{\Sigma}_0$ is composed by only one irreducible diagram. 
In the frequency space, the retarded and the Keldysh components of the self energy are given by
\begin{eqnarray}
\Sigma_{0}^R(\varepsilon) &=& \sum_k  \lambda_k^2 \left(\frac{1}{\varepsilon-\omega_k+i\eta}-\frac{1}{\varepsilon+\omega_k+i\eta}\right) \,, \\
\Sigma_{0}^K(\varepsilon) &=&   2i\mbox{Im }\Sigma_{0}^R(\varepsilon)\mathrm{coth}(\varepsilon)\, . 
\end{eqnarray}
To mimic the dissipation, the ensemble of oscillators form a bath with a continuous spectrum.
Then, by replacing the sum with an integral over the frequencies, we introduce the spectral density function for Ohmic dissipation 
\begin{equation}
J(\varepsilon) =  \sum_k \frac{\pi\lambda_k^2}{\omega_k} \left(\delta(\varepsilon-\omega_k) +\delta(\varepsilon+\omega_k)\right)= \frac{\varepsilon}{\omega} Q^{-1}\ \, .
\end{equation}
with the coefficient $Q$ corresponding to the quality factor of the oscillator. 
Finally, we can approximate $\Sigma_{0}^R(\varepsilon) \simeq \Sigma_{0}^R(\omega) $ for $Q \gg 1$ in the Dyson equation \eqref{eq:D_RA}.  
We thus obtain 
\begin{eqnarray}
\gamma_0  &=& -\mbox{Im } \Sigma_{0}^R(\omega)  = \omega/Q  \, ,\\      
\Sigma_{0}^K(\omega) &=& -2i\omega[1+2n_B(\omega)]/Q  \, .
\end{eqnarray}

\section{Retarded self-energy at zero temperature}
\label{Appendix:retardedselfenergy}
The retarded self-energy in Eq.~\eqref{eq:Sigma_RA} can be calculated analytically at zero temperature.
For completeness and comparison, we give here the expression for the real and imaginary parts. 
These expressions agree with the results of Ref. [\onlinecite{Egger:2008hh}] albeit with the generalized spin index for the spin-flip vertex interaction. 
\begin{multline}
\textrm{Re }{\Sigma^{R}_{\sigma\sigma}}(\varepsilon) 
=  \sum_{\alpha,s}\frac{\lambda^2\Gamma_\alpha^{\sigma}}{(\varepsilon-\varepsilon_{{\sigma}}-s\omega)^2+{\Gamma^{\sigma}}^2} 
\left[
\frac{\varepsilon-\varepsilon_\sigma-s\omega}{\Gamma^{\sigma}}  \right. 
\\ \left. \left(\frac{1}{2}+\frac{s}{\pi}\tan^{-1}\frac{\varepsilon_\sigma-\mu_\alpha}{\Gamma^{\sigma}}\right)+\frac{s}{\pi}\textrm{ln}\frac{\vert \varepsilon-s\omega-\mu_\alpha\vert}{\sqrt{(\varepsilon_\sigma-\mu_\alpha)^2+{\Gamma^{\sigma}}^2}} \right] \, ,
\end{multline}
and
\begin{equation}
 \textrm{Im }{\Sigma^{R}_{\sigma\sigma}} = \sum_{\alpha,s}\frac{-\lambda^2\Gamma_\alpha^\sigma\theta(s(\varepsilon-\mu_\alpha)-\omega)}{(\varepsilon-s\omega-\varepsilon_{\sigma})^2+{\Gamma^{\sigma}}^2} \, ,
\end{equation}
with $\Gamma^{\sigma} = \Gamma_l^\sigma+\Gamma_r^{\sigma}$.

\bibliography{references}

\end{document}